\documentclass[letterpaper]{article} 
\usepackage{aaai25}  
\usepackage{times}  
\usepackage{helvet}  
\usepackage{courier}  
\usepackage[hyphens]{url}  
\usepackage{graphicx} 
\usepackage{natbib}  
\usepackage{caption} 
\usepackage {pdfpages}
\usepackage{multirow}
\newtheorem{theorem}{Theorem}
\newtheorem{definition}{Definition}

\title{PCM Selector: Penalized Covariate-Mediator Selection Operator \\for Evaluating Linear Causal Effects}
\author{
    Hisayoshi Nanmo\textsuperscript{\rm 1, \rm 2}, Manabu Kuroki\textsuperscript{\rm 2}\\
}
\affiliations{
    \textsuperscript{\rm 1}Chugai Pharmaceutical Co., Ltd., Nihonbashi Muromachi, Chuo-ku, Tokyo, Japan\\
    \textsuperscript{\rm 2}Yokohama National University, Tokiwadai, Hodogaya-ku, Yokohama, Japan\\
    nanmohisayoshi@gmail.com, kuroki-manabu-zm@ynu.ac.jp

}

\begin{document}
\maketitle
\begin{abstract}
For a data-generating process for random variables that can be described with a linear structural equation model, we consider a situation in which (i) a set of covariates satisfying the back-door criterion cannot be observed or (ii) such a set can be observed, but standard statistical estimation methods cannot be applied to estimate causal effects because of multicollinearity/high-dimensional data problems. We propose a novel two-stage penalized regression approach, the penalized covariate-mediator selection operator (PCM Selector), to estimate the causal effects in such scenarios. Unlike existing penalized regression analyses, when a set of intermediate variables is available, PCM Selector provides a consistent or less biased estimator of the causal effect. In addition, PCM Selector provides a variable selection procedure for intermediate variables to obtain better estimation accuracy of the causal effects than does the back-door criterion. 
\end{abstract}
%
\begin{links}
\link{Technical Appendix}{https://doi.org/10.48550/arXiv.2412.18180}
\end{links}
\section{Introduction}
\subsection{Background}
Auxiliary variables are those that are not considered to be of interest in themselves but help us to evaluate causal effects and/or understand the data-generating process in practical studies. 
For example, an intermediate variable is often considered an auxiliary variable because it is used to evaluate causal effects \citep{Pearl2001,pearl2009}, to understand the data-generating process in the context of mediation analysis \citep{Baron1986, Imai2011, Mackinnon2008} and to improve the estimation accuracy of causal effects \citep{Cox1960, Hayashi2014}. 

In the context of linear structural equation models, this paper focuses on estimating causal effects using intermediate variables. 
For cases in which the data-generating process for random variables can be described by nonparametric structural equation models and the corresponding directed
 acyclic graph, \citet{pearl2009} provided the front-door criterion as the identification condition for causal effects based on intermediate variables. 
 In addition, in the framework of linear structural equation models,
\citet{Kuroki2000}, \citet{nanmo2021}, and \citet{kuroki2023} formulated the exact variance of causal effects based on the front-door criterion. 
Furthermore, \citet{Kuroki2004}, \citet{Hui2008}, and \citet{ramsahai2012} compared some identification conditions in terms of the asymptotic estimation accuracy of causal effects. 
On the other hand, under the assumption that a treatment variable is associated with a response variable through a univariate intermediate variable, from the viewpoint of the asymptotic estimation accuracy, \citet{Cox1960} showed that the estimation accuracy of the regression coefficient of the treatment variable on the response variable in the single linear regression model can be improved by using a joint linear regression model based on the response variable and the intermediate variable. 
In addition, 
\citet{kurokis2014} and \citet{Hayashi2014} derived the same results as \citet{Cox1960} in terms of the exact variance of causal effects. 
\citet{Gupta2021} derived the same results as \citet{kurokis2014} and \citet{Hayashi2014} for cases in which a multivariate intermediate variable is available. 

In existing studies, it is noted that causal effects can be estimated by standard statistical estimation methods, e.g., the maximum likelihood estimation (MLE) method and the ordinary least squares (OLS) method. 
Thus, many covariates affect both the treatment variable and the response variable and are highly correlated with each other in reality.
This situation leads to a multicollinearity problem, which decreases the estimation accuracy of the causal effects and leads to the formulation of an unreliable plan that prevents us from conducting appropriate policy decision-making. 
In addition, when the sample size is smaller than the number of explanatory variables in the regression analysis, 
high-dimensional data analysis also suffers from multicollinearity problems, which cause overfitting and interfere with obtaining admissible solutions for regression coefficients.
Recently, due to the development of technological advances in collecting data with many variables to better understand a given phenomenon of interest, the multicollinearity problem has become serious in many domains.
To overcome this difficulty, numerous kinds of variable selection techniques based on penalized regression analysis, e.g., the least absolute shrinkage and selection operator (LASSO),  adaptive LASSO, and Elastic Net, have been proposed by many
statistical and AI researchers and practitioners \citep{buhlmann2011, efron2004, 
Tibshirani1996, van2014, zou2006, zou2005}.
However, the present countermeasures against the multicollinearity problem are formulated independently of the problem of identifying causal effects. 
Thus, although stable results of regression analysis may be derived by these countermeasures from the viewpoint of prediction, they may yield a seriously biased
estimate of the causal effect.
\citet{nanmo2022} proposed partially adaptive L$_p$-penalized multiple regression analysis (PAL$_p$MA) based on the back-door criterion to overcome these drawbacks. 
However, because of the formulation of PAL$_p$MA, this method is not applicable to situations where a sufficient set of confounders is not available. 
In addition, PAL$_p$MA selects a set of covariates to derive a consistent or less biased estimator of causal effects but does not consider the estimation accuracy of the causal effects. 
\subsection{Contributions}
For cases in which the data-generating process for random variables can be described with a linear structural equation model, we consider a situation where (i) a set of covariates satisfying the back-door criterion cannot be observed or (ii) such a set can be observed, but standard statistical estimation methods cannot be applied to estimate causal effects because of the multicollinearity/high-dimensional data problem. 
Then, we propose a novel two-stage penalized regression approach, the penalized covariate-mediator selection operator (PCM Selector), to estimate causal effects. 
In addition to the desirable properties of PAL$_p$MA, PCM Selector also has the following properties: 
\begin{itemize}\setlength{\leftskip}{6pt}
\item[(i)] \citet{Cox1960} noted that introducing intermediate variables enables us to improve the estimation accuracy of the regression coefficients in some situations. 
However, Cox's consideration was not used in formulating
 PAL$_p$MA, LASSO, and other penalized regression analyses. 
In contrast, based on Cox's consideration, PCM Selector selects covariates and intermediate variables to evaluate the causal effects with better estimation accuracy than PAL$_p$MA and other penalized regression analyses. 
\item[(ii)] PCM Selector without intermediate variables is consistent with PAL$_p$MA. 
In this sense, PCM Selector is considered a generalization of PAL$_p$MA, and thus provides a wider class including LASSO and adaptive LASSO. 
In addition, to our knowledge, there has been much less discussion of the selection problem for intermediate variables in the context of penalized regression analysis. 
In contrast, PCM Selector selects intermediate variables in the context of penalized regression analysis.
\end{itemize}
From these properties, PCM Selector contributes to solving the multicollinearity/high-dimensional data problems of evaluating causal effects in statistical causal inference.
Given the space constraints, the proofs, several numerical experiments, and a case study are provided in the Technical Appendix.
\section{Linear Structural Causal Model}
In statistical causal inference, a directed acyclic graph (DAG) representing cause-effect relationships (data-generating process) among random variables is called a causal diagram.
A directed graph is a pair $G=(\mbox{\boldmath $V$}, \mbox{\boldmath $E$})$, where $\mbox{\boldmath $V$}$ is a finite set of vertices and the set $\mbox{\boldmath $E$}$ of directed arrows is a subset of the set $\mbox{\boldmath $V$}{\times}\mbox{\boldmath $V$}$ of ordered pairs of distinct vertices $(V_i\rightarrow V_j$ for $(V_i,V_j)\in \mbox{\boldmath $V$}{\times}\mbox{\boldmath $V$})$.
In this paper, we interchangeably refer to vertices in the DAG and random variables of the linear structural equation model.
In addition, we refer readers to \citet{pearl2009} for the graph-theoretic terminology and basic theory of structural causal models used in this paper.
\begin{definition}\label{DEFINITION 1} (Linear Structural Causal Model)
Suppose a directed acyclic graph (DAG) $G=(\mbox{\boldmath $V$}, \mbox{\boldmath $E$})$ with a set $\mbox{\boldmath $V$}=\{V_{1},V_{2},\cdots,V_{q_v}\}$ of {continuous} random variables is given.
The DAG $G$ is called a causal diagram when each child-parent family in $G$ represents a linear structural equation model
\begin{eqnarray}
V_{i}=\mu_{v_i}+\sum_{V_{j}{\in}\mbox{pa}(V_{i})}\alpha_{v_{i}v_{j}}V_{j}+\epsilon_{v_{i}},\,\,\,i=1, 2, \ldots, q_v,
\label{1}
\end{eqnarray}
where $\mbox{pa}(V_{i})$ denotes a set of parents of $V_{i}$ in DAG $G$ and random disturbances $\epsilon_{v_{1}},\epsilon_{v_{2}}, \ldots, \epsilon_{v_{q_v}}$ are assumed to be independently distributed with mean $0$ and constant variance.
In addition, $\mu_{v_i}$ is an intercept, and $\alpha_{v_{i}v_{j}}({\neq}0)$ is called a direct effect of $V_j$ on $V_i$ $(i,j=1,2,\ldots,q_v\,;\, i\neq j)$.
Then, equation (\ref{1}) is called a linear structural causal model (linear SCM) in this paper.
\end{definition}
The linear SCM is a parametric version of Pearl's nonparametric structural causal model (PCM). 

To proceed with our discussion, we define some notation.
For univariate variables 
$X$ and $Y$ and a set of variables $\mbox{\boldmath $Z$}$, let $\sigma_{xy.z}$ and $\sigma_{xx.z}$ be the conditional covariance between $X$ and $Y$ given $\mbox{\boldmath $Z$}=\mbox{\boldmath $z$}$ and the conditional variance of $X$ given $\mbox{\boldmath $Z$}=\mbox{\boldmath $z$}$, respectively. 
Then, the regression coefficient of $X$ in the single linear regression model of $Y$ on $X$ and $\mbox{\boldmath $Z$}$ is denoted by $\beta_{yx.z}={\sigma_{xy.z}}/\sigma_{xx.z}$. 
For sets of variables $\mbox{\boldmath $X$}$, $\mbox{\boldmath $Y$}$, and $\mbox{\boldmath $Z$}$ ($\mbox{\boldmath $Y$}$ can be univariate), let $\Sigma_{xy.z}$ and $\Sigma_{xx.z}$ be the conditional cross-covariance matrix between $\mbox{\boldmath $X$}$ and $\mbox{\boldmath $Y$}$ given $\mbox{\boldmath $Z$}=\mbox{\boldmath $z$}$ and the conditional variance-covariance matrix of $\mbox{\boldmath $X$}$ given $\mbox{\boldmath $Z$}=\mbox{\boldmath $z$}$, respectively.
Then, the regression coefficient vector of $\mbox{\boldmath $X$}$ in the (single/joint) linear regression model of $\mbox{\boldmath $Y$}$ on $\mbox{\boldmath $X$}$ and $\mbox{\boldmath $Z$}$ is denoted by $B_{yx.z}=\Sigma^{-1}_{xx.z}\Sigma_{xy.z}$.
In particular, for univariate $Y$ and $\mbox{\boldmath $X$}=\{X_1,X_2,...,X_{q_x}\}$,
the $i$-th element of $B_{yx.z}$ is denoted by $\beta_{yx_i.xz}$ for $i=1,2,\cdots, q_x$. 
For univariate $X$ and $\mbox{\boldmath $Y$} =\{Y_1,Y_2,...,Y_{q_y}\}$, the $i$-th element of $B_{yx.z}$ is denoted by $\beta_{y_i x.z}$ for $i=1,2,\cdots, q_y$. 
The set of variables $\mbox{\boldmath $Z$}$ is omitted from these arguments if it is an empty set. 
A similar notation is used for the remaining statistical parameters.

The main purpose of this paper is to estimate the total effects from observed data in the context of linear SCMs. 
The total effect $\tau_{yx}$ of $X$ on $Y$ is defined as the total sum of the products of the direct effects on the sequence of directed arrows along all the directed paths from $X$ to $Y$. 
To achieve our aim, we introduce the back-door and front-door-like criteria \citep{pearl2009} as the representative identification conditions for the total effects.
Here, when causal effects, such as direct, indirect, and total effects, can be determined uniquely from the variance/covariance parameters of observed variables, they are said to be identifiable; that is, they can be estimated consistently. 
Note that direct and indirect effects are also known as representative causal effects in the context of the linear SCM.
However, we are concerned with the evaluation of the total effects using intermediate variables because (i) the direct effect can be discussed in the framework of PAL$_p$MA \citep{nanmo2021} through the ``single-door criterion'' \citep{pearl2009}, and PCM Selector is a generalization of PAL$_p$MA, and (ii) the problem of evaluating the indirect effects is within the scope of PCM Selector in some situations. 
Here, the indirect effect of $X$ on $Y$ is defined as the sum of the products of the direct effects on the sequence of directed arrows along the directed paths of interest from $X$ to $Y$, excluding the direct effect of $X$ on $Y$. 
\begin{definition} (Back-Door Criterion)
Let $\{X,Y\}$ and $\mbox{\boldmath $Z$}$ be disjoint subsets of
$\mbox{\boldmath $V$}$ in DAG $G$, where $X$ is a nondescendant of $Y$.
If a set $\mbox{\boldmath $Z$}$ of vertices satisfies the following conditions relative to an ordered pair $(X, Y)$, then $\mbox{\boldmath $Z$}$ is said to satisfy the back-door criterion relative to $(X, Y)$.
\begin{itemize}\setlength{\leftskip}{6pt}
\item[(i)] No vertex in $\mbox{\boldmath $Z$}$ is a descendant of $X$; and
\item[(ii)] $\mbox{\boldmath $Z$}$ d-separates $X$ from $Y$ in the DAG obtained by deleting all the directed arrows emerging from $X$ from the DAG $G$.
\end{itemize}
\end{definition}
If a set $\mbox{\boldmath $Z$}$ of observed variables satisfies the back-door criterion relative to $(X, Y)$ in a causal diagram $G$, 
then the total effect $\tau_{yx}$ is identifiable and is given by the formula $\beta_{yx.z}$ \citep{pearl2009}. 
As seen from Rule 2 (Action/observation exchange) of do-calculus \citep{pearl2009}, note that $X$ and $Y$ of Definition 2 can be generalized to sets of variables $\mbox{\boldmath $X$}$ and $\mbox{\boldmath $Y$}$, respectively. 
Here, a covariate is defined as an element of the nondescendants of $X$ and $Y$.
In addition, a set of covariates is called a sufficient set of confounders if it satisfies the back-door criterion; otherwise, it is called an insufficient set of confounders.
\begin{definition} (Front-Door-Like Criterion)
Let $\{X,Y\}$, $\mbox{\boldmath $S$}$, $\mbox{\boldmath $Z$}_1\cup\mbox{\boldmath $Z$}_2$ be disjoint subsets of $\mbox{\boldmath $V$}$ in the DAG $G$, where $X$ is a nondescendant of $Y$.
If a set $\mbox{\boldmath $S$}$ of vertices satisfies the following conditions relative to an ordered pair $(X, Y)$ together with $\mbox{\boldmath $Z$}_1\cup\mbox{\boldmath $Z$}_{2}$, then $\mbox{\boldmath $S$}$ is said to satisfy the front-door-like criterion relative to $(X, Y)$ with $\mbox{\boldmath $Z$}_1\cup\mbox{\boldmath $Z$}_{2}$.
\begin{itemize}\setlength{\leftskip}{8pt}
\item[(i)] $\mbox{\boldmath $S$}$ intercepts all the directed paths from $X$ to $Y$;
\item[(ii)] $\mbox{\boldmath $Z$}_1$ satisfies the back-door criterion relative to $(X,\mbox{\boldmath $S$})$; and
\item[(iii)] $\mbox{\boldmath $Z$}_2\cup \{X\}$ satisfies the back-door criterion relative to $(\mbox{\boldmath $S$},Y)$.
\end{itemize}
\end{definition}
If a set $\mbox{\boldmath $S$}$ of observed variables satisfies the front-door-like criterion relative to $(X,Y)$ with $\mbox{\boldmath $Z$}_1\cup\mbox{\boldmath $Z$}_{2}$ in a causal diagram $G$, 
then the total effect $\tau_{yx}$ is identifiable and is given by the formula $B_{sx.z_1}B_{ys.xz_2}$. 
The front-door-like criterion is considered an extended version of the front-door criterion \citep{pearl2009} since it is consistent with the front-door criterion when $\mbox{\boldmath $Z$}_1\cup\mbox{\boldmath $Z$}_{2}$ is empty. 

Here, an intermediate variable relative to $(X, Y)$ is defined as one that is a descendant of $X$ and an ancestor of $Y$ simultaneously. 
In addition, a set of intermediate variables is called a sufficient set if it satisfies the front-door-like criterion; otherwise, it is called an insufficient set of intermediate variables. 
\begin{figure}[t]
\centering
\includegraphics[width=1\columnwidth]{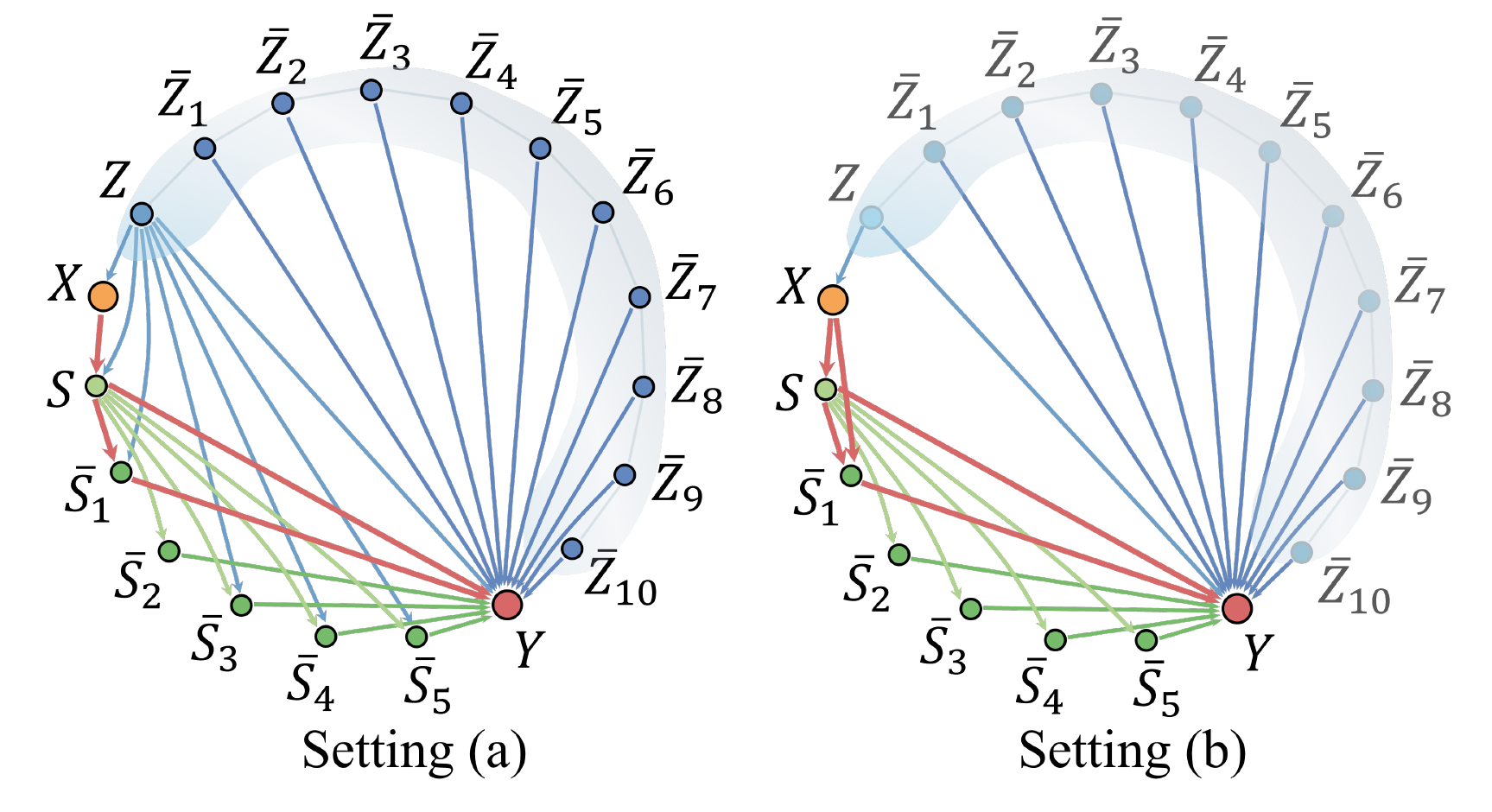}
\caption{Causal diagram. The thick red arrows show the total effect of interest. $X$: treatment variable; $Y$: response variable; $S$: intermediate variable that can be selected using prior causal knowledge; $\overline{\mbox{\boldmath $S$}}=\{\overline{S}_{1},\ldots,\overline{S}_{5}\}$: a set of intermediate variables for which it is uncertain which element should be added to evaluate the total effects; $Z$: covariate that can be selected using prior causal knowledge; $\overline{\mbox{\boldmath $Z$}}=\{\overline{Z}_{1},\ldots,\overline{Z}_{10}\}$: a set of covariates for which it is uncertain which element should be added to evaluate the total effects.}
\label{fig1}
\end{figure}
\section{PCM Selector\label{sec3}}
\subsection{Problem Setting}
In this paper, we partition a set of observed variables into the following three disjoint sets:
\begin{itemize}\setlength{\leftskip}{8pt}
\item[(i)] $\{X, Y\}$: $X$ and $Y$ are the treatment and response variables, respectively.
\item[(ii)] $\mbox{\boldmath $C$}=\mbox{\boldmath $Z$}\cup \overline{\mbox{\boldmath $Z$}}$ (`C' for covariates): a set of covariates satisfying the back-door criterion relative to $(X, Y)$ ($\mbox{\boldmath $Z$}\cap \overline{\mbox{\boldmath $Z$}}$ is empty), where $\mbox{\boldmath $Z$}$ and $\overline{\mbox{\boldmath $Z$}}$ are the first $q_z$ components and the next $q_{\overline{z}}$ components of ${\mbox{\boldmath $C$}}$, respectively.
Here, $\mbox{\boldmath $Z$}$ is a subset including some covariates selected using prior causal knowledge ($\mbox{\boldmath $Z$}$ may be an empty set, a sufficient set of confounders, or an insufficient set of confounders), but $\overline{\mbox{\boldmath $Z$}}$ is a subset of covariates for which it is uncertain which element of $\overline{\mbox{\boldmath $Z$}}$ should be added to evaluate the total effects. 
\item[(iii)] $\mbox{\boldmath $M$}=\mbox{\boldmath $S$}\cup \overline{\mbox{\boldmath $S$}}$ (`M' for intermediate variables): a set of intermediate variables satisfying the front-door-like criterion relative to $(X,Y)$ with $\mbox{\boldmath $C$}$ ($\mbox{\boldmath $S$}\cap \overline{\mbox{\boldmath $S$}}$ is empty), where $\mbox{\boldmath $S$}$ and $\overline{\mbox{\boldmath $S$}}$ are the first $q_s$ components and the next $q_{\overline{s}}$ components of $\mbox{\boldmath $M$}$, respectively. 
Here, $\mbox{\boldmath $S$}$ is a subset including some intermediate variables selected using prior causal knowledge (${\mbox{\boldmath $S$}}$ may be an empty set, a sufficient set of intermediate variables, or an insufficient set of intermediate variables), but $\overline{\mbox{\boldmath $S$}}$ is a subset for which it is uncertain which element of $\overline{\mbox{\boldmath $S$}}$ should be added to evaluate the total effects. 
\end{itemize}
Then, for sample size $n$, consider the following joint linear regression model of $\{Y\}\cup\mbox{\boldmath $M$}$:
\begin{eqnarray}
&\mbox{\boldmath $y$}=\mbox{\boldmath $x$}{\beta}_{yx.cm}+\mbox{\boldmath $c$}B_{yc.xm}+\mbox{\boldmath $m$}B_{ym.xc}+\mbox{\boldmath $\epsilon$}_{y.xcm},&\\
&\mbox{\boldmath $m$}=\mbox{\boldmath $x$}B_{mx.c}+\mbox{\boldmath $c$}B_{mc.x}+\mbox{\boldmath $\epsilon$}_{m.xc},\label{e3}&
\end{eqnarray}
where $\mbox{\boldmath $x$}$ and $\mbox{\boldmath $y$}$ represent $n$-dimensional observation vectors of $X$ and $Y$, respectively.
$\mbox{\boldmath $c$}$ and $\mbox{\boldmath $m$}$ are an $n\times (q_z+q_{\overline{z}})$ observation matrix of $\mbox{\boldmath $C$}$ and an $n\times (q_s+q_{\overline{s}})$ observation matrix of $\mbox{\boldmath $M$}$, respectively. 
Here, $\mbox{\boldmath $x$}$, $\mbox{\boldmath $y$}$, $\mbox{\boldmath $c$}$ and $\mbox{\boldmath $m$}$ are standardized to sample mean $0$ and sample variance $1$ in advance.
In addition, we assume that the elements of the random error vector $\mbox{\boldmath $\epsilon$}_{y.xcm}$ are independent and identically distributed with mean $0$ and finite variance {$\sigma_{yy.xcm}$}. 
Furthermore, the column vectors of the random error matrix $\mbox{\boldmath $\epsilon$}_{m.xc}$ are independent and identically distributed with zero mean vector and variance-covariance matrix $\Sigma_{mm.xc}$ for $M\in\mbox{\boldmath $M$}$ and are also independent of the elements of $\mbox{\boldmath $\epsilon$}_{y.xcm}$.  

Under the above setting, this paper focuses on situations where the sum-of-squares matrix of $\{X\}\cup\mbox{\boldmath $S$}\cup\mbox{\boldmath $Z$}$ is invertible but that of $\{X\}\cup\mbox{\boldmath $C$}\cup\mbox{\boldmath $M$}$ is not; this is because if it is invertible, then the total effect is estimable by the OLS method \citep{pearl2009}.

\subsection{Estimator}

For univariates $X$ and $Y$ and a set of variables $\mbox{\boldmath $Z$}$, let $s_{xx.z}$ and $s_{xy.z}$ be the sum-of-squares of $X$ given $\mbox{\boldmath $Z$}$ and the sum of cross-products between $X$ and $Y$ given $\mbox{\boldmath $Z$}$, respectively.
In addition, for sets of variables $\mbox{\boldmath $X$}$, $\mbox{\boldmath $Y$}$, and $\mbox{\boldmath $Z$}$ ($\mbox{\boldmath $Y$}$ can be univariate), let $S_{xx.z}$ and $S_{xy.z}$ be the sum-of-squares matrix of $\mbox{\boldmath $X$}$ given $\mbox{\boldmath $Z$}$ and the sum-of-cross-products matrix between $\mbox{\boldmath $X$}$ and $\mbox{\boldmath $Y$}$ given $\mbox{\boldmath $Z$}$, respectively. 
Here, the set of variables $\mbox{\boldmath $Z$}$ is omitted from these arguments if it is an empty set. 
A similar notation is used for the remaining sums of squares/cross-products.
Furthermore, $\mbox{\boldmath $0$}_q$, $\mbox{\boldmath $0$}_{q,r}$, $\mbox{\boldmath $1$}_q$ and $I_{q}$ are a $q$-dimensional zero vector, a $q\times r$ zero matrix, a $q$-dimensional one vector, and a $q\times q$ identity matrix, respectively. 

Then, the proposed penalized regression approach, PCM Selector, is formulated as follows:

First, when the sum-of-squares matrix of $\{X\}\cup\mbox{\boldmath $C$}\cup\mbox{\boldmath $M$}$ is invertible, let 
\begin{eqnarray}
\begin{array}{c}
\hspace{-1mm}\hat{\beta}_{yx.cm}=s_{xy.cm}/s_{xx.cm},\ \ 
\hat{B}_{yc.xm}=S^{-1}_{cc.xm}S_{cy.xm},\\
\hat{B}_{ym.xc}=S^{-1}_{mm.xc}S_{my.xc},
\end{array}
\label{3}
\end{eqnarray}
and when the sum-of-squares matrix of $\{X\}\cup\mbox{\boldmath $C$}\cup\mbox{\boldmath $M$}$ is not invertible, let
\begin{eqnarray}
&&\hspace{-5mm}\left(\tilde{\beta}_{yx.cm},\tilde{B}_{ys.xc\overline{s}},\tilde{B}_{yz.xm\overline{z}},\tilde{B}_{y\overline{s}.xcs},\tilde{B}_{y\overline{z}.xmz}\right)^{T}\nonumber\\
&&\hspace{-7mm}=\hspace{-1mm}\left(
\begin{array}{ccccc}\hspace{-1.2mm}
n\lambda+s_{xx}&\hspace{-1mm}S_{xs}&\hspace{-1mm}S_{xz}&\hspace{-1mm}S_{x\overline{s}}&\hspace{-1mm}S_{x\overline{z}}\\
S_{sx}&\hspace{-1mm}S_{ss}&\hspace{-1mm}S_{sz}&\hspace{-1mm}S_{s\overline{s}}&\hspace{-1mm}S_{s\overline{z}}\\
S_{zx}&\hspace{-1mm}S_{zs}&\hspace{-1mm}S_{zz}&\hspace{-1mm}S_{z\overline{s}}&\hspace{-1mm}S_{z\overline{z}}\\
S_{\overline{s}x}&\hspace{-1mm}S_{\overline{s}s}&\hspace{-1mm}S_{\overline{s}z}&\hspace{-1mm}n\lambda I_{q_{\overline{s}}}+S_{\overline{s}\overline{s}}&\hspace{-1mm}S_{\overline{s}\overline{z}}\\
S_{\overline{z}x}&\hspace{-1mm}S_{\overline{z}s}&\hspace{-1mm}S_{\overline{z}z}&\hspace{-1mm}S_{\overline{z}\overline{s}}&\hspace{-1mm}n\lambda I_{q_{\overline{z}}}+S_{\overline{z}\overline{z}}
\end{array}\hspace{-1.2mm}\right)^{\hspace{-0.5mm}-1}\nonumber\\
&&\hspace{27mm}\times\left(s_{xy},S_{sy},S_{zy},S_{\overline{s}y},S_{\overline{z}y}\right)^{T}\label{4}
\end{eqnarray}
for the  penalty parameter $\lambda>0$, where $S_{yx.z}=S^T_{xy.z}$ and the superscript ``$T$'' represents the transposed vector/matrix. 
Here, equation (\ref{4}) is consistent with equation (\ref{3}) for $\lambda=0$. In addition, when the sum-of-squares matrix of $\{X\}\cup\mbox{\boldmath $C$}$ is not invertible, let
\begin{eqnarray}
\left(\hspace{-1.2mm}
\begin{array}{c}
\tilde{B}_{mx.c}\\
\tilde{B}_{mz.x\overline{z}}\\
\tilde{B}_{m\overline{z}.xz}
\end{array}
\hspace{-1.2mm}\right)\hspace{-1.2mm}
=\hspace{-1mm}\left(
\begin{array}{ccc}
s_{xx}&\hspace{-1mm}S_{xz}&\hspace{-1mm}S_{x\overline{z}}\\
S_{zx}&\hspace{-1mm}S_{zz}&\hspace{-1mm}S_{z\overline{z}}\\
S_{\overline{z}x}&\hspace{-1mm}S_{\overline{z}z}&\hspace{-1mm}n\rho I_{q_{\overline{z}}}+S_{\overline{z}\overline{z}}
\end{array}\hspace{-1.2mm}\right)^{\hspace{-0.5mm}-1}\hspace{-2mm}
\left(\hspace{-1.2mm}
\begin{array}{c}
S_{xm}\\ 
S_{zm}\\ 
S_{\overline{z}m}\\
\end{array}\hspace{-1.2mm}
\right)\label{1000}
\end{eqnarray}
for the penalty parameter $\rho>0$. 
For $p=1,2$, consider the $L_p$-penalized loss function
\begin{eqnarray}
\lefteqn{L_p({\beta}_{yx.cm},B_{yc.xm},B_{ym.xc})} \nonumber \\
 &&\hspace{-5mm}=\frac{1}{2n}\|\mbox{\boldmath $y$}-\mbox{\boldmath $x$}{\beta}_{yx.cm}-\mbox{\boldmath $c$}B_{yc.xm}-\mbox{\boldmath $m$}B_{ym.xc}\|^2_2 \nonumber \\
&&+\lambda_{p}\left(\zeta_{p}\|\beta_{yx.cm}\|_{p}^{p}+\xi_{p}\|\mbox{\boldmath $\gamma$}_{\overline{s}x.c}\odot B_{y\overline{s}.xcs}\|_{p}^{p}\right.\nonumber \\
&&\left.\hspace{4mm}+(1-\zeta_{p}-\xi_{p})\|\mbox{\boldmath $\gamma$}_{y\overline{z}.xmz}\odot B_{y\overline{z}.xmz}\|_{p}^{p}\right)\label{e5}
\end{eqnarray}
for the tuning parameters $\zeta_{p}\geq 0$ and $\xi_{p}\geq 0$ such that $\zeta_{p}+\xi_{p}\in[0,1]$, the penalty parameter $\lambda_p$ corresponding to the $L_p$ norm ($\lambda_p\geq 0$), and the multivariate response type $L_p$-penalized loss function
\begin{eqnarray}
\lefteqn{\hspace{-4mm}L_p(B_{mx.c},B_{mc.x})=\frac{1}{2n}\|\mbox{\boldmath $m$}-\mbox{\boldmath $x$}B_{mx.c}-\mbox{\boldmath $c$}B_{mc.x}\|^2_F} \nonumber \\
&&\hspace{17mm}+\rho_{p}\|\mathrm{vec}\left(\mbox{\boldmath $\gamma$}_{m\overline{z}.xz}\odot B_{m\overline{z}.xz}\right)\|_{p}^{p}\label{e10}
\end{eqnarray}
for the penalty parameter $\rho_p$ corresponding to the $L_p$ norm ($\rho_p\geq 0$). 
Here, $\odot$, $\|\cdot\|^p_p$, and $\|\cdot\|_{F}$ refer to the Hadamard product, the $L_p$ norm, and the Frobenius norm, respectively. 
In addition, for $\overline{s}_{i}\in\overline{\mbox{\boldmath $S$}}\, (i=1,2,...,q_{\overline{s}})$, $\overline{z}_{i}\in\overline{\mbox{\boldmath $Z$}}\, (i=1,2,...,q_{\overline{z}})$ and $m_{i}\in{\mbox{\boldmath $M$}}\, (i=1,2,...,q_{m})$, 
the standardized weight vectors $\mbox{\boldmath $\gamma$}_{\overline{s}x.c}$ and $\mbox{\boldmath $\gamma$}_{y\overline{z}.xmz}$ and the standardized weight matrix $\mbox{\boldmath $\gamma$}_{m\overline{z}.xz}$ 
are given by
\begin{eqnarray}
&&\displaystyle\hspace{-7mm}\lefteqn{\mbox{\boldmath $\gamma$}_{\overline{s}x.c}=\left(\sum_{i=1}^{q_{\overline{s}}}\frac{1}{|\tilde{\beta}_{\overline{s}_{i}x.c}|}\right)^{-1}}\nonumber\\
&&\hspace{13mm}\times\left(\frac{1}{|\tilde{\beta}_{\overline{s}_{1}x.c}|},\frac{1}{|\tilde{\beta}_{\overline{s}_{2}x.c}|},\ldots,\frac{1}{|\tilde{\beta}_{\overline{s}_{q_{\overline{s}}}x.c}|}\right)^{T},\label{weight1}\\
&&\displaystyle\lefteqn{\hspace{-7mm}\mbox{\boldmath $\gamma$}_{y\overline{z}.xmz}=\left(\sum_{i=1}^{q_{\overline{z}}}\frac{1}{|\tilde{\beta}_{y\overline{z}_{i}.xmc}|}\right)^{-1}}\nonumber\\
&&\hspace{1mm}\times\left(\frac{1}{|\tilde{\beta}_{y\overline{z}_{1}.xmc}|},\frac{1}{|\tilde{\beta}_{y\overline{z}_{2}.xmc}|},\ldots,\frac{1}{|\tilde{\beta}_{y\overline{z}_{q_{\overline{z}}}.xmc}|}\right)^{T},\label{weight2}\\
&&\displaystyle\hspace{-7mm}\mbox{\boldmath $\gamma$}_{m\overline{z}.xz}\hspace{-1mm}=\hspace{-1.2mm}\left[\hspace{-1mm}\left(\sum_{k=1}^{q_{\overline{z}}}\sum_{\ell=1}^{q_{m}}\frac{1}{|\tilde{\beta}_{m_{\ell}\overline{z}_{k}.xc}|}\hspace{-0.6mm}\right)^{\hspace{-1.1mm}-1}\hspace{-3.9mm}\frac{1}{|\tilde{\beta}_{m_{j}\overline{z}_{i}.xc}|}\hspace{-0.6mm}\right]_{\hspace{-0.9mm}1\le i\le q_{\overline{z}},1\le j\le q_{m}}\label{weight3}
\end{eqnarray}
respectively, where $|\cdot |$ refers to the absolute value, and the vec operator, $\mathrm{vec}(A)$, denotes the vectorization of an $q\times r$ matrix $A$, which is the $q\times r$-dimensional vector obtained by stacking the columns of matrix $A$ on top of one another. 
Equation (\ref{e5}) is different from the standard penalized loss function in the following ways: 
\begin{itemize}\setlength{\leftskip}{8pt}
\item[(i)] The penalty parameter $\lambda_p$ is not assigned to $B_{yz.xm\overline{z}}$ and $B_{ys.xc\overline{s}}$ in equation (\ref{e5}) in order not to remove covariates ($\mbox{\boldmath $Z$}$) and intermediate variables ($\mbox{\boldmath $S$}$) selected using prior causal knowledge.
\item[(ii)] The weight vector constructed by $\tilde{B}_{\overline{s}x.c}$ of $\tilde{B}_{mx.c}=(\tilde{B}_{{s}x.c},\tilde{B}_{\overline{s}x.c})$, but not that constructed by $\tilde{B}_{y\overline{s}.xcs}$, is assigned to $B_{y\overline{s}.xcs}$. Equation (\ref{weight1}) shows that the indirect effect of $X$ on $Y$ decreases 
via $\overline{S}_i\in \overline{\mbox{\boldmath $S$}}$ to zero when $B_{\overline{s}_ix.c}$ approaches zero. 
\item[(iii)] Standardizing each weight vector enable us to fairly select covariates and intermediate variables in order of priority.
\end{itemize}
For $p=1$, ${\beta}_{yx.cm}$, $B_{yc.xm}$ and $B_{ym.xc}$, which minimize equation (\ref{e5}), and $B_{mx.c}$ and $B_{mc.x}$, which minimize equation (\ref{e10}), are called PCM estimators, denoted by $\check{\beta}_{yx.cm}^{\dagger}$, $\check{B}_{yc.xm}^{\dagger}$, $\check{B}_{ym.xc}^{\dagger}$, $\check{B}_{mx.c}^{\dagger}$, and $\check{B}_{mc.x}^{\dagger}$, respectively.
Since equation (\ref{e5}) is consistent with the partially adaptive $L_p$-penalized loss function given by \citet{nanmo2022} when $\zeta_{p}$ and $\xi_{p}$ respectively are zero and $\mbox{\boldmath $M$}$ is an empty set, PCM Selector is considered a generalization of PAL$_p$MA. 
Under the assumption that the sum-of-squares matrix of $\{X\}\cup\mbox{\boldmath $C$}\cup\mbox{\boldmath $M$}$ is invertible, letting $\lambda_p=0$,
$\beta_{yx.cm}$, $B_{yc.xm}$ and $B_{ym.xc}$, which minimize equation (\ref{e5}), are given by the OLS estimators, i.e., equation (\ref{3}).
In addition, 
Let $p=2$, {$\lambda_2=3\lambda>0$}, $\zeta_{p}=1/3$, $\xi_{p}=1/3$, $\mbox{\boldmath $\gamma$}_{\overline{s}x.c}=\mbox{\boldmath $1$}_{q_{\overline{s}}}$ and $\mbox{\boldmath $\gamma$}_{y\overline{z}.xmz}=\mbox{\boldmath $1$}_{q_{\overline{z}}}$. 
Then, ${\beta}_{yx.cm}$, $B_{ym.xc}$ and $B_{yc.xm}$, which minimize equation (\ref{e5}), are given by the ridge-type estimators in equation (\ref{4}).

Here, in order to avoid confusion by the notation in the following discussion, regarding equations (\ref{e5}) and (\ref{e10}) for $p=1$, let $\{X\}$, $\overline{\mbox{\boldmath $S$}}$ and $\overline{\mbox{\boldmath $Z$}}$ be active sets for a given $\lambda_1,\rho_1>0$, which is a subset of variables with nonzero regression coefficients that do not include any elements of $\mbox{\boldmath $Z$}\cup\mbox{\boldmath $S$}$ 
.
In addition, let $q_{\overline{s}}$ and $q_{\overline{z}}$ be the numbers of variables in the active sets $\overline{\mbox{\boldmath $S$}}$ and $\overline{\mbox{\boldmath $Z$}}$, respectively. 
Then, under the assumption that the sum-of-squares matrix of explanatory variables $\{X\}\cup\mbox{\boldmath $C$}\cup\mbox{\boldmath $M$}$ is invertible,
when $X$ is active, $\check{\beta}_{yx.cm}^{\dagger}$, $\check{B}_{ys.xc\overline{s}}^{\dagger}$, $\check{B}_{y\overline{s}.xcs}^{\dagger}$ and $\check{B}^{\dagger}_{mx.c}$ are given by
\begin{eqnarray}
\lefteqn{\hspace{-6mm}\left(\check{\beta}_{yx.cm}^{\dagger},\check{B}_{ys.xc\overline{s}}^{\dagger},\check{B}_{y\overline{s}.xcs}^{\dagger}\right)^{T}\hspace{-1mm}=\left(\hat{\beta}_{yx.cm},\hat{B}_{ys.xc\overline{s}},\hat{B}_{y\overline{s}.xcs}\right)^{T}}\nonumber\\
&&\hspace{10mm}+n\lambda_{1}\left(
\begin{array}{ccc}
-1&\hat{B}_{\overline{s}x.sc}&\hat{B}_{\overline{z}x.zm}\\
\hat{B}_{xs.c\overline{s}}&\hat{B}_{\overline{s}s.xc}&\hat{B}_{\overline{z}s.xz\overline{s}}\\
\hat{B}_{x\overline{s}.cs}&-I_{q_{\overline{s}}}&\hat{B}_{\overline{z}\overline{s}.xsz}
\end{array}\right)\nonumber\\
&&\hspace{-12mm}\times\hspace{-1mm}\left(\hspace{-1.7mm}
\begin{array}{c}
\zeta_{1}s_{xx.cm}^{-1}\mathrm{sign}(\check{\beta}_{yx.cm}^{\dagger})\\
\xi_{1}S_{\overline{s}\overline{s}.xcs}^{-1}\mbox{\boldmath $\gamma$}_{\overline{s}x.c}\hspace{-1mm}\odot\mathrm{sign}(\check{B}_{y\overline{s}.xcs}^{\dagger})\\
(1-\zeta_{1}-\xi_{1})S_{\overline{z}\overline{z}.xmz}^{-1}\mbox{\boldmath $\gamma$}_{y\overline{z}.xmz}\hspace{-1mm}\odot\mathrm{sign}(\check{B}_{y\overline{z}.xmz}^{\dagger})
\end{array}\hspace{-1.7mm}\right)\hspace{-1mm},
\label{e8}
\end{eqnarray}
\begin{eqnarray}
\lefteqn{\check{B}_{mx.c}^{\dagger}=\hat{B}_{mx.c}}\nonumber\\
&&+n\rho_{1}\hat{B}_{\overline{z}x.z}S_{\overline{z}\overline{z}.xz}^{-1}\mbox{\boldmath $\gamma$}_{m\overline{z}.xz}\odot\mathrm{sign}(\check{B}_{m\overline{z}.xz}^{\dagger})\label{mxc},
\end{eqnarray}
where 
\begin{equation}
\hspace{-2mm}\begin{array}{l}
\hat{B}_{\overline{s}x.sc}=s^{-1}_{xx.sc}S_{x\overline{s}.sc},\quad\hat{B}_{\overline{z}x.zm}=s^{-1}_{xx.zm}S_{x\overline{z}.zm},\\
\hat{B}_{xs.c\overline{s}}=S^{-1}_{ss.c\overline{s}}S_{sx.x\overline{s}},\quad
\hat{B}_{\overline{s}s.xc}=S^{-1}_{ss.xc}S_{s\overline{s}.xc},\\
\hat{B}_{\overline{z}s.xz\overline{s}}=S^{-1}_{ss.xz\overline{s}}S_{s\overline{z}.xz\overline{s}},\quad\hat{B}_{x\overline{s}.cs}=S^{-1}_{\overline{s}\overline{s}.cs}S_{\overline{s}x.cs},
\\
\hat{B}_{\overline{z}\overline{s}.xsz}=S^{-1}_{\overline{s}\overline{s}.xsz}S_{\overline{s}\overline{z}.xsz}, \quad\hat{B}_{ys.xc\overline{s}}=S^{-1}_{ss.xz\overline{s}}S_{ys.xz\overline{s}},\\
\hat{B}_{y\overline{s}.xcs}=S^{-1}_{\overline{s}\overline{s}.xzs}S_{y\overline{s}.xzs}
\end{array}\label{e9}
\end{equation}
In addition, for a $q\times r$ matrix $A=(a_{ij})_{1\le i\le q, 1\le j\le r}$, $\mbox{sign}(A)=(\mbox{sign}(a_{ij}))_{1\le i\le q, 1\le j\le r}$, where
\begin{eqnarray}
\mbox{sign} (a_{ij})=\left\{\begin{array}{cl}
1&a_{ij}>0\\
0&a_{ij}=0\\
-1&a_{ij}<0
\end{array}\right.
\label{9}
\end{eqnarray}
for $i=1,2,...,q$, $j=1,2,...,r$. When $X$ is not active, $\check{\beta}_{yx.cm}^{\dagger}$ is evaluated as zero. 
In addition, $\check{B}_{ys.c\overline{s}}^{\dagger}$ and $\check{B}_{y\overline{s}.cs}^{\dagger}$ are obtained by omitting the subscript $x$ in equation (\ref{e8}) except for $\mbox{\boldmath $\gamma$}_{\overline{s}x.c}$ and replacing $\hat{B}_{\overline{s}x.sc}$, $\hat{B}_{\overline{z}x.zm}$, $\hat{B}_{xs.c\overline{s}}$, $\hat{B}_{x\overline{s}.cs}$ and $s^{-1}_{xx.cm}$ with zeros in equation (\ref{e8}). 
Note that $\mbox{\boldmath $\gamma$}_{\overline{s}x.c}$ is given by equation (\ref{weight1}) regardless of whether $X$ is active or not.

Here, for $\lambda_{2}, \rho_2, \rho'_2 \geq 0$ and $\xi_2\in[0,1]$, to reduce the bias, based on the derived active sets, the following estimators are considered:
\begin{itemize}\setlength{\leftskip}{4pt}
\item[\bf(a)]{\bf$\tilde{B}^{\dagger}_{xc.m}$ and $\tilde{B}^{\dagger}_{xm.c}$:} $B_{xc.m}$ and $B_{xm.c}$ that minimize
\end{itemize}
\begin{eqnarray}
\lefteqn{L_{2}(
B_{xc.m},B_{xm.c})}\nonumber\\
&&\hspace{-5mm}=\frac{1}{2n}\|\mbox{\boldmath $x$}-\mbox{\boldmath $z$}B_{xz.\overline{z}m}-\overline{\mbox{\boldmath $z$}}B_{x\overline{z}.zm}-\mbox{\boldmath $s$}B_{xs.c\overline{s}}-\overline{\mbox{\boldmath $s$}}B_{x\overline{s}.cs}\|^2_2\nonumber\\
&&\hspace{7mm}+\lambda_{2}\left\{\xi_{2}\|B_{x\overline{s}.cs}\|_{2}^{2}+(1-\xi_{2})\|B_{x\overline{z}.zm}\|_{2}^{2}\right\},
\end{eqnarray}
\begin{itemize}\setlength{\leftskip}{4pt}
\item[\bf(b)]{\bf $\tilde{B}^{\dagger}_{\overline{s}x.cs}$, $\tilde{B}^{\dagger}_{\overline{s}s.xc}$ and $\tilde{B}^{\dagger}_{\overline{s}c.xs}$:} $B_{\overline{s}x.cs}$, $B_{\overline{s}s.xc}$ and $B_{\overline{s}c.xs}$ that minimize
\end{itemize}
\begin{eqnarray}
\lefteqn{L_{2}(
B_{\overline{s}x.cs}, B_{\overline{s}s.xc},B_{\overline{s}c.xs})}\nonumber\\
&&\hspace{-5mm}=\frac{1}{2n}\|\overline{s}-\mbox{\boldmath $x$}B_{\overline{s}x.sc}-\mbox{\boldmath $s$}B_{\overline{s}s.xc}-\mbox{\boldmath $z$}B_{\overline{s}z.xs\overline{z}}-\overline{\mbox{\boldmath $z$}}B_{\overline{s}\overline{z}.xsz}\|^2_F\nonumber\\
&&\hspace{35mm}+\rho_{2}\|\mathrm{vec}\left(B_{\overline{s}\overline{z}.xsz}\right)\|_{2}^{2},
\end{eqnarray}
\begin{itemize}\setlength{\leftskip}{4pt}
\item[\bf(c)]{\bf$\tilde{B}^{\dagger}_{\overline{z}x.zm}$, $\tilde{B}^{\dagger}_{\overline{z}z.xm}$ and $\tilde{B}^{\dagger}_{\overline{z}m.xz}$:} $B_{\overline{z}x.zm}$, $B_{\overline{z}z.xm}$ and $B_{\overline{z}m.xz}$ that minimize
\end{itemize}
\begin{eqnarray}
\lefteqn{L_{2}(
B_{\overline{z}x.zm}, B_{\overline{z}z.xm},B_{\overline{z}m.xz})}\nonumber\\
&&\hspace{-6mm}=\hspace{-1mm}\frac{1}{2n}\hspace{-0.1mm}\|\hspace{-0.1mm}\overline{\mbox{\boldmath $z$}}\hspace{-0.1mm}-\hspace{-0.1mm}\mbox{\boldmath $x$}B_{\overline{z}x.zm}\hspace{-0.1mm}-\hspace{-0.1mm}\mbox{\boldmath $z$}B_{\overline{z}z.xm}\hspace{-0.1mm}-\hspace{-0.1mm}\mbox{\boldmath $s$}B_{\overline{z}s.xz\overline{s}}\hspace{-0.1mm}-\hspace{-0.1mm}\overline{\mbox{\boldmath $s$}}B_{\overline{z}\overline{s}.xzs}\|^{\hspace{-0.1mm}2}_{\hspace{-0.1mm}F}\nonumber\\
&&\hspace{35mm}+{\rho'_{2}}\|\mathrm{vec}\left(B_{\overline{z}\overline{s}.xsz}\right)\|_{2}^{2}.
\end{eqnarray}
Then, based on equations (\ref{e8}) and (\ref{mxc}), when $X$ is active, consider
\begin{eqnarray}
\lefteqn{\hspace{-6mm}\left(\check{\beta}_{yx.cm}^{*},\check{B}_{ys.xc\overline{s}}^{*},\check{B}_{y\overline{s}.xcs}^{*}\right)^{T}=\left(\check{\beta}_{yx.cm}^{\dagger},\check{B}_{ys.xc\overline{s}}^{\dagger},\check{B}_{y\overline{s}.xcs}^{\dagger}\right)^{T}}\nonumber\\
&&\hspace{10mm}-n\lambda_{1}\left(
\begin{array}{ccc}
-1&\tilde{B}_{\overline{s}x.sc}^{\dagger}&\tilde{B}_{\overline{z}x.zm}^{\dagger}\\
\tilde{B}_{xs.c\overline{s}}^{\dagger}&\tilde{B}_{\overline{s}s.xc}^{\dagger}&\tilde{B}_{\overline{z}s.xz\overline{s}}^{\dagger}\\
\tilde{B}_{x\overline{s}.cs}^{\dagger}&-I_{q_{\overline{s}}}&\tilde{B}_{\overline{z}\overline{s}.xsz}^{\dagger}
\end{array}\right)\nonumber\\
&&\hspace{-12mm}\times\hspace{-1mm}\left(\hspace{-1.7mm}
\begin{array}{c}
\zeta_{1}\tilde{s}_{xx.cm}^{\dagger -1}\mathrm{sign}(\check{\beta}_{yx.cm}^{\dagger})\\
\xi_{1}\tilde{S}_{\overline{s}\overline{s}.xcs}^{\dagger+}\mbox{\boldmath $\gamma$}_{\overline{s}x.c}\hspace{-1mm}\odot\mathrm{sign}(\check{B}_{y\overline{s}.xcs}^{\dagger})\\
(1-\zeta_{1}-\xi_{1})\tilde{S}_{\overline{z}\overline{z}.xmz}^{\dagger+}\mbox{\boldmath $\gamma$}_{y\overline{z}.xmz}\hspace{-1mm}\odot \mathrm{sign}(\check{B}_{y\overline{z}.xmz}^{\dagger})
\end{array}\hspace{-1.7mm}\right)\hspace{-1mm},\label{correct1}
\end{eqnarray}
\begin{eqnarray}
\lefteqn{\check{B}_{mx.c}^{*}=\check{B}_{mx.c}^{\dagger}}\nonumber\\
&&-n{\rho_{1}}\hat{B}_{\overline{z}x.z}\hat{S}_{\overline{z}\overline{z}.xz}^{+}\mbox{\boldmath $\gamma$}_{m\overline{z}.xz}\odot\mathrm{sign}(\check{B}_{m\overline{z}.xz}^{\dagger}),
\end{eqnarray}
where $\mbox{\boldmath $m$}$ and $\mbox{\boldmath $c$}$ of $\check{B}_{mx.c}^{*}$ are constructed by both $\mbox{\boldmath $S$}\cup \mbox{\boldmath $Z$}$ and a subset of $\overline{\mbox{\boldmath $S$}}\cup \overline{\mbox{\boldmath $Z$}}$ corresponding to the active sets of $\check{B}_{y\overline{s}.xcs}^{\dagger}$ and $\check{B}_{y\overline{c}.xmz}^{\dagger}$, 
\begin{eqnarray}
&&\hspace{-10mm}\tilde{s}^{\dagger}_{xx.cm}
=\|\mbox{\boldmath $x$}-\mbox{\boldmath $c$}\tilde{B}^{\dagger}_{xc.m}-\mbox{\boldmath $m$}\tilde{B}^{\dagger}_{xm.c}\|_{2}^{2},\\
&&\hspace{-10mm}\tilde{S}^{\dagger}_{\overline{s}\overline{s}.xcs}
=||\mbox{\boldmath $\overline{s}$}-\mbox{\boldmath $x$}\tilde{B}^{\dagger}_{\overline{s}x.cs}-\mbox{\boldmath $s$}\tilde{B}^{\dagger}_{\overline{s}s.xc}-\mbox{\boldmath $c$}\tilde{B}^{\dagger}_{\overline{s}c.xs}||_G,\\
&&\hspace{-10mm}\tilde{S}^{\dagger}_{\overline{z}\overline{z}.xmz}\hspace{-1mm}
=||\mbox{\boldmath $\overline{z}$}-\mbox{\boldmath $x$}\tilde{B}^{\dagger}_{\overline{z}x.cs}\hspace{-0.5mm}-\mbox{\boldmath $m$}\tilde{B}^{\dagger}_{\overline{z}m.xz}\hspace{-0.5mm}-\mbox{\boldmath $z$}\tilde{B}^{\dagger}_{\overline{z}z.xm}||_G,\\
&&\hspace{-10mm}\hat{S}_{\overline{z}\overline{z}.xz}
=||\mbox{\boldmath $\overline{z}$}-\mbox{\boldmath $x$}\hat{B}_{\overline{z}x.z}-\mbox{\boldmath $z$}\hat{B}_{\overline{z}z.x}||_G,
\end{eqnarray}
and $\|A\|_{G}$ and $A^{+}$ denote the gram matrix $A^{T}A$ and the generalized inverse of a matrix $A$ \citep{bernstein2009}, respectively. 
{When $X$ is not active, $\check{\beta}^*_{yx.cm}$ is evaluated as zero. 
In addition, $\check{B}^*_{ys.c\overline{s}}$ and $\check{B}^*_{y\overline{s}.cs}$ are obtained by omitting the subscript $x$ from equation (\ref{correct1}) except for $\mbox{\boldmath $\gamma$}_{\overline{s}x.c}$ and replacing $\tilde{B}_{\overline{s}x.sc}^{\dagger}$, $\tilde{B}_{\overline{z}x.zm}^{\dagger}$, $\tilde{B}^{\dagger}_{xs.c\overline{s}}$, $\tilde{B}^{\dagger}_{x\overline{s}.cs}$ and $\tilde{s}^{\dagger -1}_{xx.cm}$ with zeros in equation (\ref{correct1}). 
Note that $\mbox{\boldmath $\gamma$}_{\overline{s}x.c}$ is given by equation (\ref{weight1}) regardless of whether $X$ is active or not. }

Then, we formulate the modified PCM estimator of the total effect ${\tau}_{yx}$ as 
\begin{eqnarray*}
\check{\tau}_{yx}^{*}=\check{\beta}_{yx.cm}^{*}+\check{B}_{mx.c}^{*}\check{B}_{ym.xc}^{*}
\end{eqnarray*}
when $X$ is active according to equation (\ref{e5}) and
\begin{eqnarray*}
\check{\tau}_{yx}^{*}=\check{B}_{mx.c}^{*}\check{B}_{ym.c}^{*}
\end{eqnarray*}
when $X$ is not active according to equation (\ref{e5}). 
Hereafter, the modified PCM estimator is called the PCM estimator.

Regarding PCM estimators, the following theorems hold:
\begin{theorem}\label{THEOREM 1}
For an active set $\mbox{\boldmath $M$}\cup \mbox{\boldmath $C$}$, when the OLS estimators are available, if $X$ is conditionally independent of $Y$ given $\mbox{\boldmath $M$}\cup \mbox{\boldmath $C$}$, then the following inequalities approximately hold under the normality:
\begin{eqnarray}
&&\hspace*{-15mm}\mbox{var}(\check{B}_{mx.c}^{*}\check{B}_{ym.c}^{*})
\leq \mbox{var}(\hat{B}_{mx\cdot c}\hat{B}_{ym\cdot c})
\leq \mbox{var}(\hat{\beta}_{yx\cdot c})\\
&&\mbox{var}(\check{B}_{mx.c}^{*}\check{B}_{ym.c}^{*})
\leq \mbox{var}(\check{\beta}_{yx.c}^{*})
\end{eqnarray}
for the optimal tuning and penalty parameters. 
\end{theorem}
The first inequality is given in the Technical Appendix. 
The second inequality is shown in \citet{kurokis2014,kuroki2016}. 
Theorem 1 shows that the estimation accuracy of the total effect can be improved compared to that of the OLS method through PCM Selector based on a set of variables that make $X$ and $Y$ conditionally independent. 
\begin{theorem}\label{THEOREM 2}
For an active set $\mbox{\boldmath $M$}\cup \mbox{\boldmath $C$}$, when the OLS estimators are available, 
if $X$ is conditionally independent of $Y$ given $\mbox{\boldmath $M$}\cup \mbox{\boldmath $C$}$ and $\mbox{\boldmath $M$}'\cup \mbox{\boldmath $C$}$, the following inequalities approximately hold under the normality: 
\begin{eqnarray}
\lefteqn{\hspace{-7mm}\mbox{var}(\check{B}_{m'x.c}^{*}\check{B}_{ym'.c}^{*})
\leq \mbox{var}(\hat{B}_{m'x.c}\hat{B}_{ym'.c})}\nonumber\\
&&\hspace{20mm}\leq \mbox{var}(\hat{B}_{mx.c}\hat{B}_{ym.c})
\end{eqnarray}
for $\mbox{\boldmath $M$}'\subset \mbox{\boldmath $M$}$.  
\end{theorem}
The first inequality is simply obtained from Theorem 1, and the second inequality is shown in \citet{kurokis2014,kuroki2016}. 
Theorem 2 provides a statistical guideline for selecting a set of intermediate variables to derive a more efficient estimator of the total effects.

\section{Numerical Experiment\label{sec4}}
\begin{table*}[!ht]
\begin{center}
{\small \begin{tabular}{ccccccccccccc} \\
\cline{1-13}
\multirow{2}{*}{Setting (a)}&  \multicolumn{4}{c}{$\tau_{yx}=0.045$}& &\multicolumn{7}{c}{parameter settings}\\\cline{2-5}\cline{7-13}
& Mean & SD & Bias&Sign&& $\lambda$& $\eta$& $\phi$& $\lambda_{1}$& $\rho_{1}$& $\zeta_{1}$&$\xi_{1}$\\\cline{1-13}
LASSO & 0.013 & {0.045} & -0.033 &0.117 && 0.407& -& -& -& -& -& -\\
adaptive LASSO & 0.017 & 0.057 & -0.028 &0.138 && 0.407& 0.100& -& -& -& -& -\\
Elastic Net & 0.017 & 0.054 & -0.028&0.156 && 0.399& -& 0.910& -& -& -& -\\
PAL$_1$MA & 0.054 & 0.792 & {0.009} &0.528 && 0.294& 1.200& -& -& -& -& - \\
PCM Selector & 0.036 & 0.718 & -0.010&0.526 & & -& -& -& 0.017& 0.213& 0.270& 0.190 \\
Front-door-like (including $x$) & -0.008 & 1.577 & -0.053 &0.515 && -& -& -& -& -& -& -\\
Front-door-like (not including $x$) & 0.030 & 1.051 & -0.015 &0.524 && -& -& -& -& -& -& - \\
Back-door & 0.054 & 1.591 & 0.009&0.532 & & -& -& -& -& -& -& -\\\cline{1-13}\\
\end{tabular}
\begin{tabular}{cccccccccc}
\cline{1-10}
\multirow{2}{*}{Setting (b)}&  \multicolumn{4}{c}{$\tau_{yx}=0.402$}& &\multicolumn{4}{c}{parameter settings}\\\cline{2-5}\cline{7-10}
& Mean & SD & Bias&Sign& & $\lambda_{1}$& $\rho_{1}$& $\zeta_{1}$&$\xi_{1}$\\\cline{1-10}
PCM Selector & 0.448  & {0.549}  & {0.046} &0.808& &0.346&-&0.000&1.000 \\
Front-door (minimal) & 0.468 & 0.552 & 0.066 &0.818& &-&-&-&- \\
Front-door (whole) & 0.462  & 0.692  & 0.060&0.770& &-&-&-&- \\\cline{1-10}
\end{tabular}}
\end{center}
\caption{Results based on cross-validation. Mean: sample mean; SD: standard deviation; Bias: bias between the true value and the sample mean; Sign: coincidence rate between the signs of the true value and the estimates; Front-door-like (including $x$): the treatment variable $X$, the intermediate variable $S$ and the set of covariates $\mbox{\boldmath $C$}$ are used for the front-door-like criterion; Front-door-like (not including $x$): the intermediate variable $S$ and the set of covariates $\mbox{\boldmath $C$}$ are used for the front-door-like criterion; Back-door: the set of covariates $\mbox{\boldmath $C$}$ is used for the back-door criterion; Front-door (minimal): a minimally sufficient set of intermediate variables is used for the front-door criterion. Front-door (whole): the set of intermediate variables $\mbox{\boldmath $M$}$ is used for the front-door criterion.$\lambda, \lambda_{1}, \rho_{1}$: penalty parameters; $\eta$: tuning parameter for the adaptive weights \citep{zou2006}; $\phi$: tuning parameter for the elastic net penalty \citep{zou2005}; $\zeta_{1}, \xi_{1}$: tuning parameters; $\lambda=3.157$, $\rho=69.484$ for equations (\ref{4}) and (\ref{1000}) in Setting (a) and $\lambda=3.726$ for equation (\ref{1000}) in Setting (b); $\tau_{yx}$: true value of total effect.}
\end{table*}
In this section, we present a numerical experiment to compare the performances of LASSO, adaptive LASSO, Elastic Net, PAL$_1$MA, the OLS method, the two-stage least squares (TSLS) method and PCM Selector. For brevity, consider the linear SCM
\begin{equation}
\left.
\begin{array}{l}
Y=\alpha_{ys}S+\alpha_{yz}Z+\overline{\mbox{\boldmath $S$}}A_{y\overline{s}}+\overline{\mbox{\boldmath $Z$}}A_{y\overline{z}}+\epsilon_{y}\\
\overline{\mbox{\boldmath $S$}}=XA_{\overline{s}x}+SA_{\overline{s}s}+ZA_{\overline{s}z}+\mbox{\boldmath $\epsilon$}_{\overline{s}}\\
S=\alpha_{sx}X+\alpha_{sz}Z+\epsilon_{s}\\
X=\alpha_{xz}Z+\epsilon_{x}
\end{array}\right\}
\label{aaaaa}
\end{equation}
for Figure~1, where $\overline{\mbox{\boldmath $Z$}}$ and $\overline{\mbox{\boldmath $S$}}$ include {10} covariates and {5} intermediate variables ($\mbox{\boldmath $M$}=\{S\}\cup\overline{\mbox{\boldmath $S$}}$), respectively. 
In Figure~1, Setting (a) shows that (i) $S$ satisfies the front-door-like criterion relative to $(X,Y)$ with $Z$ and (ii) $Z$ satisfies the back-door criterion relative to $(X,Y)$ and Setting (b) shows that (i) $\{S,\overline{S}_{1}\}$ satisfies the front-door criterion relative to $(X,Y)$ and (ii) $\mbox{\boldmath $C$}=\{Z,\overline{\mbox{\boldmath $Z$}}\}$ satisfies the back-door criterion relative to $(X,Y)$ but is unobserved. Here, $S$ and $\{S,\overline{S}_{1}\}$ are the minimally sufficient sets of intermediate variables that satisfies the front-door-like criterion for Setting (a), and satisfies the front-door criterion for Setting (b), respectively.

To set up the numerical experiment, we first construct the population variance-covariance matrix.
To eliminate arbitrariness, the true values of the direct effects are $\alpha_{ys}=0.4$, $\alpha_{\overline{s}s}=0.2\ (\in A_{\overline{s}s})$, $\alpha_{\overline{s}_{2}x}, \alpha_{\overline{s}_{3}x}, \alpha_{\overline{s}_{4}x}, \alpha_{\overline{s}_{5}x}\ (\in A_{\overline{s}x})$ are set to $0$ and $\alpha_{y\overline{z}_{1}},\alpha_{y\overline{z}_{2}},\ldots,\alpha_{y\overline{z}_{10}}(\in A_{y\overline{z}})$, $\alpha_{y\overline{s}_{2}},\alpha_{y\overline{s}_{3}},\alpha_{y\overline{s}_{4}},\alpha_{y\overline{s}_{5}}(\in A_{y\overline{s}})$ are randomly and independently generated according to a uniform distribution on the interval $[-0.2,0.2]$ in the both settings (a) and (b). The other direct effects are given as follows: Setting (a) $\alpha_{xz}=0.8$, $\alpha_{\overline{s}_{1}x}=0.0$,  $\alpha_{sx}=0.1$, $\alpha_{yz}=\alpha_{sz}=\alpha_{\overline{s}z}=0.2\ (\alpha_{\overline{s}z}\in A_{\overline{s}z})$, $\alpha_{y\overline{s}_{1}}$ is randomly generated according to a uniform distribution on the interval $[-0.2,0.2]$; Setting (b) $\alpha_{xz}=\alpha_{\overline{s}_{1}x}=\alpha_{y\overline{s}_{1}}=0.2$, $\alpha_{sx}=0.8$,   $\alpha_{yz}=\alpha_{sz}=\alpha_{\overline{s}z}=0.0\ (\alpha_{\overline{s}z}\in A_{\overline{s}z})$.

In addition, we assume that the random disturbances $\epsilon_{x}$, $\epsilon_{y}$, $\epsilon_{s}$ and $\mbox{\boldmath $\epsilon$}_{\overline{s}}$ independently follow a normal distribution in which $X$, $Y$, $S$, $\overline{\mbox{\boldmath $S$}}$ and $\mbox{\boldmath $C$}$ are standardized to mean $0$ and the unit variance. Furthermore, the population variance-covariance matrix of $\mbox{\boldmath $C$}$ is randomly determined according to \citet{Pourahmadi15}.

We generated 15 random samples of {18} variables from
a multivariate normal distribution with a zero mean vector and the above variance-covariance matrix for 5000 replications.
Table 1 shows the basic statistics of the total effects estimated by LASSO, adaptive LASSO, Elastic Net, PAL$_1$MA, the OLS method, the TSLS methods, and PCM Selector based on the given penalty and tuning parameters. Here, the TSLS methods are based on front-door-like criterion in Setting (a) and based on front-door criterion in Setting (b). In addition, for the OLS and TSLS methods, we select a set of covariates $\mbox{\boldmath $C$}$ in Setting (a).  In Setting (b), it is assumed that a set of covariates is not observed, and thus the total effect can not be estimated by using the back-door criterion. Regarding the parameter tuning for LASSO, adaptive LASSO, Elastic Net, PAL$_1$MA and PCM Selector, see Section C in the Technical Appendix. 

According to Table 1, PCM Selector provides better estimation accuracy than PAL$_1$MA and the least squares methods. In addition, Table 1 shows that PCM Selector generally provides an estimation that is biased but less biased than the TSLS methods in the present parameter setting. Furthermore, the coincidence rates between the signs of the estimated total effects and the true total effects are low for LASSO, adaptive LASSO, and Elastic Net. This would be serious because it provides a misleading interpretation that the external intervention of the treatment variable $X$ does not have no effect on the change of $Y$. In contrast, the coincidence rates of PCM Selector and PAL$_1$MA are not low.

The Technical Appendix provides further discussion.

\section{Conclusion}
In current situations where advanced artificial intelligence technology enables us to collect large datasets, it is not difficult to observe many covariates and intermediate variables.
In such situations, it would be reasonable to consider such sets of variables to evaluate total effects.
However, it is difficult to evaluate the total effects reliably when multicollinearity/high-dimensional data problems occur in this situation.  
To solve this problem, we establish PCM Selector, which is considered as a wider class, including adaptive LASSO and PAL$_p$MA, to provide a less biased estimator of total effects with better estimation accuracy. 
In addition, through numerical experiments and a case study in the Technical Appendix, we confirmed that PCM Selector is superior to other methods. 
Interestingly, there are some situations where the total effect is not identifiable, but the indirect effects are identifiable \citep{inoue2022}. 
Although the current penalized regression analyses are not applicable to such situations, PCM Selector is applicable for evaluating the indirect effect.

Finally, although PCM Selector is formulated based on single/joint linear regression models,
it would be interesting to extend our approach to a wide variety of statistical models, including generalized linear models. 
Such an extension would be straightforward - the loss function would be replaced with a more general form. 
This extension will be left for future work.
\begin{figure}[h]
\centering
\includegraphics[width=0.98\columnwidth]{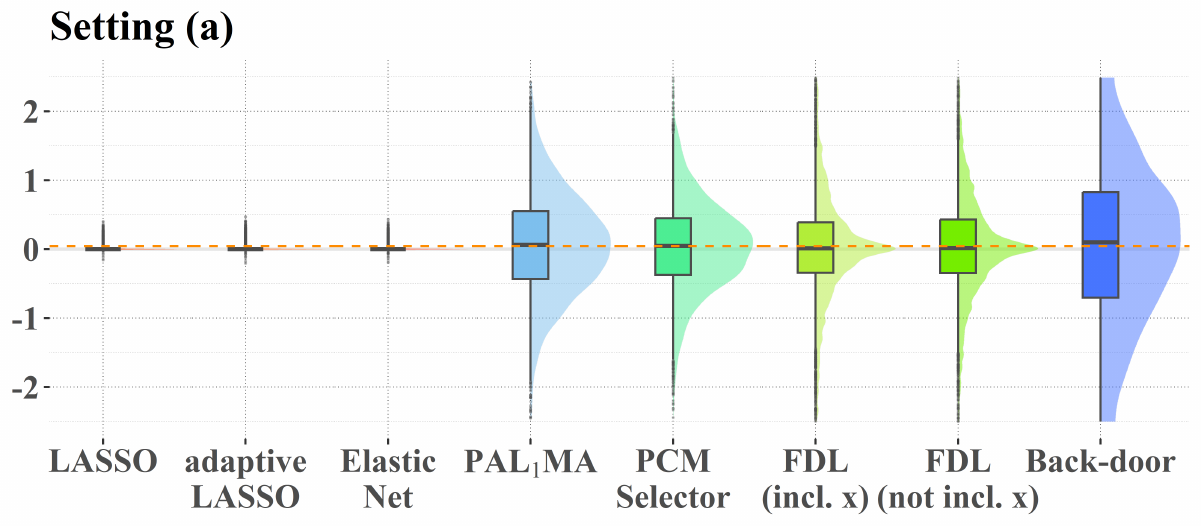}

\includegraphics[width=0.98\columnwidth]{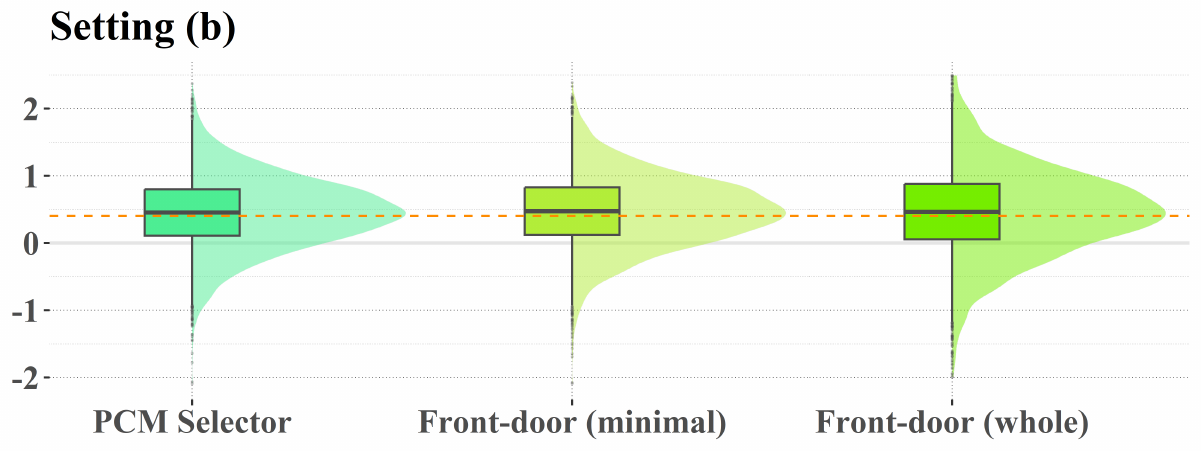}

\caption{Violin plots of estimated total effects. The dashed lines show the true total effects. FDL: Front-door-like.}
\label{fig2}
\end{figure}
\section{Acknowledgements}
We would like to acknowledge the helpful comments of the five anonymous reviewers. This research was partially funded by the JSPS Grant-in-Aids for Scientific Research 23K21700 and 24K14851.

\bibliography{AAAI2025-main}


\section*{Supplementary material}

# PCM Selector: Penalized Covariate-Mediator Selection Operator for Evaluating Linear Causal Effects
## - Technical Appendix-

Hisayoshi Nanmo[1, 2], Manabu Kuroki [2]
[1]Chugai Pharmaceutical Co., Ltd., Nihonbashi Muromachi, Chuo-ku, Tokyo, Japan
[2]Yokohama National University, Tokiwadai, Hodogaya-ku, Yokohama, Japan
nanmohisayoshi@gmail.com, kuroki-manabu-zm@ynu.ac.jp

## A Derivation of PCM estimator

Note that PCM estimator can be derived by repeated application of the blockwise inversion formula of the invertible matrix (Bernstein, 2009): For the invertible matrix

$$\begin{pmatrix} A & B \\ C & D \end{pmatrix}, \tag{A.1}$$

we have

$$\begin{pmatrix} A & B \\ C & D \end{pmatrix}^{-1} = \begin{pmatrix} A^{-1} + A^{-1}B\left(D - CA^{-1}B\right)^{-1}CA^{-1} & -A^{-1}B\left(D - CA^{-1}B\right)^{-1} \\ -\left(D - CA^{-1}B\right)^{-1}CA^{-1} & \left(D - CA^{-1}B\right)^{-1} \end{pmatrix}$$

$$= \begin{pmatrix} \left(A - BD^{-1}C\right)^{-1} & -\left(A - BD^{-1}C\right)^{-1}BD^{-1} \\ -D^{-1}C\left(A - BD^{-1}C\right)^{-1} & D^{-1} + D^{-1}C\left(A - BD^{-1}C\right)^{-1}BD^{-1} \end{pmatrix}, \tag{A.2}$$

where $A$, $D$, $D - CA^{-1}B$ and $A - BD^{-1}C$ are invertible square submatrices. Then, the derivation of PCM estimators is based on the following steps:

**Step 1:** The derivation of $\check{B}^\dagger_{mx.c}$

**Step 2:** The derivation of $\check{\beta}^\dagger_{yx.cm}$, $\check{B}^\dagger_{ys.xc\overline{s}}$ and $\check{B}^\dagger_{y\overline{s}.xcs}$

### Step 1: The derivation of $\check{B}^\dagger_{mx.c}$

When the sum-of-squares matrix of $\{X\} \cup \boldsymbol{C}$ is invertible, by using the idea of the sub-derivative, for $p = 1$, we find that the values of $B_{mx.c}$, $B_{mz.x\overline{z}}$ and $B_{m\overline{z}.xz}$ that minimize equation (8) are given by

$$\begin{pmatrix} \check{B}^\dagger_{mx.c} \\ \check{B}^\dagger_{mz.x\overline{z}} \\ \check{B}^\dagger_{m\overline{z}.xz} \end{pmatrix} = \begin{pmatrix} \hat{B}_{mx.c} \\ \hat{B}_{mz.x\overline{z}} \\ \hat{B}_{m\overline{z}.xz} \end{pmatrix} - n\rho_1 \begin{pmatrix} S_{xx} & S_{xz} & S_{x\overline{z}} \\ S_{zx} & S_{zz} & S_{z\overline{z}} \\ S_{\overline{z}x} & S_{\overline{z}z} & S_{\overline{z}\overline{z}} \end{pmatrix}^{-1} \begin{pmatrix} \mathbf{0}^T_{q_m} \\ \mathbf{0}_{q_z,q_m} \\ \boldsymbol{\gamma}_{m\overline{z}.xz} \odot \mathrm{sign}(\check{B}^\dagger_{m\overline{z}.xz}) \end{pmatrix}. \tag{A.3}$$

Then, letting

$$\begin{pmatrix} S_{xx} & S_{xz} & S_{x\overline{z}} \\ S_{zx} & S_{zz} & S_{z\overline{z}} \\ S_{\overline{z}x} & S_{\overline{z}z} & S_{\overline{z}\overline{z}} \end{pmatrix}^{-1} = \begin{pmatrix} S^{xx} & S^{xz} & S^{x\overline{z}} \\ S^{zx} & S^{zz} & S^{z\overline{z}} \\ S^{\overline{z}x} & S^{\overline{z}z} & S^{\overline{z}\overline{z}} \end{pmatrix}, \tag{A.4}$$

since we have

$$\begin{pmatrix} S^{x\overline{z}} \\ S^{z\overline{z}} \end{pmatrix} = \begin{pmatrix} -\hat{B}_{\overline{z}x.z}S^{-1}_{\overline{z}\overline{z}.xz} \\ -\hat{B}_{\overline{z}z.x}S^{-1}_{\overline{z}\overline{z}.xz} \end{pmatrix}, \tag{A.5}$$

from the blockwise inversion formula of the invertible matrix (Bernstein, 2009), we derive

$$\begin{aligned} \check{B}^\dagger_{mx.c} &= \hat{B}_{mx.c} - n\rho_1 S^{x\overline{z}} \boldsymbol{\gamma}_{m\overline{z}.xz} \odot \mathrm{sign}(\check{B}^\dagger_{m\overline{z}.xz}) \\ &= \hat{B}_{mx.c} + n\rho_1 \hat{B}_{\overline{z}x.z} S^{-1}_{\overline{z}\overline{z}.xz} \boldsymbol{\gamma}_{m\overline{z}.xz} \odot \mathrm{sign}(\check{B}^\dagger_{m\overline{z}.xz}). \end{aligned} \tag{A.6}$$

1

**Step 2: The derivation of $\check{\beta}_{yx.cm}^{\dagger}$, $\check{B}_{ys.xc\bar{s}}^{\dagger}$ and $\check{B}_{y\bar{s}.xcs}^{\dagger}$**

Similarly, when the sum-of-squares matrix of $\{X\} \cup \boldsymbol{C} \cup \boldsymbol{M}$ is invertible, by using the idea of the subderivative, for $p = 1$, we find that the values of $\beta_{yx.cm}$, $B_{ys.xc\bar{s}}$, $B_{yz.xm\bar{z}}$, $B_{y\bar{s}.xcs}$ and $B_{y\bar{z}.xmz}$ that minimize equation (7) are given by

$$\begin{pmatrix} \check{\beta}_{yx.cm}^{\dagger} \\ \check{B}_{ys.xc\bar{s}}^{\dagger} \\ \check{B}_{yz.xm\bar{z}}^{\dagger} \\ \check{B}_{y\bar{s}.xcs}^{\dagger} \\ \check{B}_{y\bar{z}.xmz}^{\dagger} \end{pmatrix} = \begin{pmatrix} \hat{\beta}_{yx.cm} \\ \hat{B}_{ys.xc\bar{s}} \\ \hat{B}_{yz.xm\bar{z}} \\ \hat{B}_{y\bar{s}.xcs} \\ \hat{B}_{y\bar{z}.xmz} \end{pmatrix} - n\lambda_1 \begin{pmatrix} s_{xx} & S_{xs} & S_{xz} & S_{x\bar{s}} & S_{x\bar{z}} \\ S_{sx} & S_{ss} & S_{sz} & S_{s\bar{s}} & S_{s\bar{z}} \\ S_{zx} & S_{zs} & S_{zz} & S_{z\bar{s}} & S_{z\bar{z}} \\ S_{\bar{s}x} & S_{\bar{s}s} & S_{\bar{s}z} & S_{\bar{s}\bar{s}} & S_{\bar{s}\bar{z}} \\ S_{\bar{z}x} & S_{\bar{z}s} & S_{\bar{z}z} & S_{\bar{z}\bar{s}} & S_{\bar{z}\bar{z}} \end{pmatrix}^{-1}$$

$$\times \begin{pmatrix} \zeta_1 \text{sign}(\check{\beta}_{yx.cm}^{\dagger}) \\ \boldsymbol{0}_{q_s} \\ \boldsymbol{0}_{q_z} \\ \xi_1 \boldsymbol{\gamma}_{\bar{s}x.c} \odot \text{sign}(\check{B}_{y\bar{s}.xcs}^{\dagger}) \\ (1 - \zeta_1 - \xi_1) \boldsymbol{\gamma}_{y\bar{z}.xmz} \odot \text{sign}(\check{B}_{y\bar{z}.xmz}^{\dagger}) \end{pmatrix}. \quad (A.7)$$

Then, letting

$$\begin{pmatrix} s_{xx} & S_{xs} & S_{xz} & S_{x\bar{s}} & S_{x\bar{z}} \\ S_{sx} & S_{ss} & S_{sz} & S_{s\bar{s}} & S_{s\bar{z}} \\ S_{zx} & S_{zs} & S_{zz} & S_{z\bar{s}} & S_{z\bar{z}} \\ S_{\bar{s}x} & S_{\bar{s}s} & S_{\bar{s}z} & S_{\bar{s}\bar{s}} & S_{\bar{s}\bar{z}} \\ S_{\bar{z}x} & S_{\bar{z}s} & S_{\bar{z}z} & S_{\bar{z}\bar{s}} & S_{\bar{z}\bar{z}} \end{pmatrix}^{-1} = \begin{pmatrix} s^{xx} & S^{xs} & S^{xz} & S^{x\bar{s}} & S^{x\bar{z}} \\ S^{sx} & S^{ss} & S^{sz} & S^{s\bar{s}} & S^{s\bar{z}} \\ S^{zx} & S^{zs} & S^{zz} & S^{z\bar{s}} & S^{z\bar{z}} \\ S^{\bar{s}x} & S^{\bar{s}s} & S^{\bar{s}z} & S^{\bar{s}\bar{s}} & S^{\bar{s}\bar{z}} \\ S^{\bar{z}x} & S^{\bar{z}s} & S^{\bar{z}z} & S^{\bar{z}\bar{s}} & S^{\bar{z}\bar{z}} \end{pmatrix}, \quad (A.8)$$

we have

$$\begin{pmatrix} s^{xx} & S^{xs} \\ S^{sx} & S^{ss} \end{pmatrix} = \begin{pmatrix} s_{xx.cm}^{-1} & -s_{xx.c\bar{s}}^{-1} S_{xs.c\bar{s}} S_{ss.xc\bar{s}}^{-1} \\ -S_{ss.c\bar{s}}^{-1} S_{sx.c\bar{s}} s_{xx.cm}^{-1} & S_{ss.xc\bar{s}}^{-1} \end{pmatrix}$$

$$= \begin{pmatrix} s_{xx.cm}^{-1} & -\hat{B}_{sx.c\bar{s}} S_{ss.xc\bar{s}}^{-1} \\ -\hat{B}_{xs.c\bar{s}} s_{xx.cm}^{-1} & S_{ss.xc\bar{s}}^{-1} \end{pmatrix}, \quad (A.9)$$

$$\begin{pmatrix} S^{\bar{s}\bar{s}} & S^{\bar{s}\bar{z}} \\ S^{\bar{z}\bar{s}} & S^{\bar{z}\bar{z}} \end{pmatrix} = \begin{pmatrix} S_{\bar{s}\bar{s}.xcs}^{-1} & -S_{\bar{s}\bar{s}.xsz}^{-1} S_{\bar{s}\bar{z}.xsz} S_{\bar{z}\bar{z}.xmz}^{-1} \\ -S_{\bar{z}\bar{z}.xsz}^{-1} S_{\bar{z}\bar{s}.xsz} S_{\bar{s}\bar{s}.xcs}^{-1} & S_{\bar{z}\bar{z}.xmz}^{-1} \end{pmatrix}$$

$$= \begin{pmatrix} S_{\bar{s}\bar{s}.xcs}^{-1} & -\hat{B}_{\bar{z}\bar{s}.xsz} S_{\bar{z}\bar{z}.xmz}^{-1} \\ -\hat{B}_{\bar{s}\bar{z}.xsz} S_{\bar{s}\bar{s}.xcs}^{-1} & S_{\bar{z}\bar{z}.xmz}^{-1} \end{pmatrix}, \quad (A.10)$$

$$\begin{pmatrix} S^{zx} & S^{zs} \\ S^{\bar{s}x} & S^{\bar{s}s} \\ S^{\bar{z}x} & S^{\bar{z}s} \end{pmatrix} = - \begin{pmatrix} \hat{B}_{xz.\bar{s}\bar{z}} & \hat{B}_{sz.\bar{s}\bar{z}} \\ \hat{B}_{x\bar{s}.c} & \hat{B}_{s\bar{s}.c} \\ \hat{B}_{x\bar{z}.z\bar{s}} & \hat{B}_{s\bar{z}.z\bar{s}} \end{pmatrix} \begin{pmatrix} s_{xx.cm}^{-1} & -\hat{B}_{sx.c\bar{s}} S_{ss.xc\bar{s}}^{-1} \\ -\hat{B}_{xs.c\bar{s}} s_{xx.cm}^{-1} & S_{ss.xc\bar{s}}^{-1} \end{pmatrix} \quad (A.11)$$

$$= \begin{pmatrix} \hat{B}_{sz.\bar{s}\bar{z}} \hat{B}_{xs.c\bar{s}} s_{xx.cm}^{-1} - \hat{B}_{xz.\bar{s}\bar{z}} s_{xx.cm}^{-1} & \hat{B}_{xz.\bar{s}\bar{z}} \hat{B}_{sx.c\bar{s}} S_{ss.xc\bar{s}}^{-1} - \hat{B}_{sz.\bar{s}\bar{z}} S_{ss.xc\bar{s}}^{-1} \\ \hat{B}_{s\bar{s}.c} \hat{B}_{xs.c\bar{s}} s_{xx.cm}^{-1} - \hat{B}_{x\bar{s}.c} s_{xx.cm}^{-1} & \hat{B}_{x\bar{s}.c} \hat{B}_{sx.c\bar{s}} S_{ss.xc\bar{s}}^{-1} - \hat{B}_{s\bar{s}.c} S_{ss.xc\bar{s}}^{-1} \\ \hat{B}_{s\bar{z}.z\bar{s}} \hat{B}_{xs.c\bar{s}} s_{xx.cm}^{-1} - \hat{B}_{x\bar{z}.z\bar{s}} s_{xx.cm}^{-1} & \hat{B}_{x\bar{z}.z\bar{s}} \hat{B}_{sx.c\bar{s}} S_{ss.xc\bar{s}}^{-1} - \hat{B}_{s\bar{z}.z\bar{s}} S_{ss.xc\bar{s}}^{-1} \end{pmatrix},$$

$$\begin{pmatrix} S^{x\bar{s}} & S^{x\bar{z}} \\ S^{s\bar{s}} & S^{s\bar{z}} \\ S^{z\bar{s}} & S^{z\bar{z}} \end{pmatrix} = - \begin{pmatrix} \hat{B}_{\bar{s}x.sz} & \hat{B}_{\bar{z}x.sz} \\ \hat{B}_{\bar{s}s.xz} & \hat{B}_{\bar{z}s.xz} \\ \hat{B}_{\bar{s}z.xs} & \hat{B}_{\bar{z}z.xs} \end{pmatrix} \begin{pmatrix} S_{\bar{s}\bar{s}.xcs}^{-1} & -\hat{B}_{\bar{z}\bar{s}.xsz} S_{\bar{z}\bar{z}.xmz}^{-1} \\ -\hat{B}_{\bar{s}\bar{z}.xsz} S_{\bar{s}\bar{s}.xcs}^{-1} & S_{\bar{z}\bar{z}.xmz}^{-1} \end{pmatrix} \quad (A.12)$$

$$= \begin{pmatrix} \hat{B}_{\bar{z}x.sz} \hat{B}_{\bar{s}\bar{z}.xsz} S_{\bar{s}\bar{s}.xcs}^{-1} - \hat{B}_{\bar{s}x.sz} S_{\bar{s}\bar{s}.xcs}^{-1} & \hat{B}_{\bar{s}x.sz} \hat{B}_{\bar{z}\bar{s}.xsz} S_{\bar{z}\bar{z}.xmz}^{-1} - \hat{B}_{\bar{z}x.sz} S_{\bar{z}\bar{z}.xmz}^{-1} \\ \hat{B}_{\bar{z}s.xz} \hat{B}_{\bar{s}\bar{z}.xsz} S_{\bar{s}\bar{s}.xcs}^{-1} - \hat{B}_{\bar{s}s.xz} S_{\bar{s}\bar{s}.xcs}^{-1} & \hat{B}_{\bar{s}s.xz} \hat{B}_{\bar{z}\bar{s}.xsz} S_{\bar{z}\bar{z}.xmz}^{-1} - \hat{B}_{\bar{z}s.xz} S_{\bar{z}\bar{z}.xmz}^{-1} \\ \hat{B}_{\bar{z}z.xs} \hat{B}_{\bar{s}\bar{z}.xsz} S_{\bar{s}\bar{s}.xcs}^{-1} - \hat{B}_{\bar{s}z.xs} S_{\bar{s}\bar{s}.xcs}^{-1} & \hat{B}_{\bar{s}z.xs} \hat{B}_{\bar{z}\bar{s}.xsz} S_{\bar{z}\bar{z}.xmz}^{-1} - \hat{B}_{\bar{z}z.xs} S_{\bar{z}\bar{z}.xmz}^{-1} \end{pmatrix}$$

from the blockwise inversion formula of the invertible matrix (Bernstein, 2009). Thus, we derive

$$\check{\beta}_{yx.cm}^{\dagger} = \hat{\beta}_{yx.cm} - n\lambda_1 \left\{ \zeta_1 s^{xx} \text{sign}(\check{\beta}_{yx.cm}^{\dagger}) + \xi_1 S^{x\bar{s}} \boldsymbol{\gamma}_{\bar{s}x.c} \odot \text{sign}(\check{B}_{y\bar{s}.xcs}^{\dagger}) \right.$$

$$+(1-\zeta_1-\xi_1)S^{x\bar{z}}\gamma_{y\bar{z}.xmz}\odot\text{sign}(\check{B}^\dagger_{y\bar{z}.xmz})\Big\}$$
$$=\hat{\beta}_{yx.cm}-n\lambda_1\Big\{\zeta_1 s^{-1}_{xx.cm}\text{sign}(\check{\beta}^\dagger_{yx.cm})-\xi_1\hat{B}_{\bar{s}x.sc}S^{-1}_{\bar{s}\bar{s}.xcs}\gamma_{\bar{s}x.c}\odot\text{sign}(\check{B}^\dagger_{y\bar{s}.xcs})$$
$$-(1-\zeta_1-\xi_1)\hat{B}_{\bar{z}x.zm}S^{-1}_{\bar{z}\bar{z}.xmz}\gamma_{y\bar{z}.xmz}\odot\text{sign}(\check{B}^\dagger_{y\bar{z}.xmz})\Big\}, \qquad (A.13)$$

$$\check{B}^\dagger_{ys.xc\bar{s}} = \hat{B}_{ys.xc\bar{s}}-n\lambda_1\Big\{\zeta_1 S^{sx}\text{sign}(\check{\beta}^\dagger_{yx.cm})+\xi_1 S^{s\bar{s}}\gamma_{\bar{s}x.c}\odot\text{sign}(\check{B}^\dagger_{y\bar{s}.xcs})$$
$$+(1-\zeta_1-\xi_1)S^{s\bar{z}}\gamma_{y\bar{z}.xmz}\odot\text{sign}(\check{B}^\dagger_{y\bar{z}.xmz})\Big\}$$
$$=\hat{B}_{ys.xc\bar{s}}-n\lambda_1\Big\{-\zeta_1\hat{B}_{xs.c\bar{s}}s^{-1}_{xx.cm}\text{sign}(\check{\beta}^\dagger_{yx.cm})$$
$$-\xi_1\hat{B}_{\bar{s}s.xc}S^{-1}_{\bar{s}\bar{s}.xcs}\gamma_{\bar{s}x.c}\odot\text{sign}(\check{B}^\dagger_{y\bar{s}.xcs})$$
$$-(1-\zeta_1-\xi_1)\hat{B}_{\bar{z}s.xz\bar{s}}S^{-1}_{\bar{z}\bar{z}.xmz}\gamma_{y\bar{z}.xmz}\odot\text{sign}(\check{B}^\dagger_{y\bar{z}.xmz})\Big\}, \qquad (A.14)$$

$$\check{B}^\dagger_{y\bar{s}.xcs} = \hat{B}_{y\bar{s}.xcs}-n\lambda_1\Big\{\zeta_1 S^{\bar{s}x}\text{sign}(\check{\beta}^\dagger_{yx.cm})+\xi_1 S^{\bar{s}s}\gamma_{\bar{s}x.c}\odot\text{sign}(\check{B}^\dagger_{y\bar{s}.xcs})$$
$$+(1-\zeta_1-\xi_1)S^{\bar{s}z}\gamma_{y\bar{z}.xmz}\odot\text{sign}(\check{B}^\dagger_{y\bar{z}.xmz})\Big\}$$
$$=\hat{B}_{y\bar{s}.xcs}-n\lambda_1\Big\{-\zeta_1\hat{B}_{x\bar{s}.cs}s^{-1}_{xx.cm}\text{sign}(\check{\beta}^\dagger_{yx.cm})+\xi_1 S^{-1}_{\bar{s}\bar{s}.xcs}\gamma_{\bar{s}x.c}\odot\text{sign}(\check{B}^\dagger_{y\bar{s}.xcs})$$
$$-(1-\zeta_1-\xi_1)\hat{B}_{\bar{z}\bar{s}.xsz}S^{-1}_{\bar{z}\bar{z}.xmz}\gamma_{y\bar{z}.xmz}\odot\text{sign}(\check{B}^\dagger_{y\bar{z}.xmz})\Big\}. \qquad (A.15)$$

By combining the above equations, we derive

$$\begin{pmatrix}\check{\beta}^\dagger_{yx.cm}\\ \check{B}^\dagger_{ys.xc\bar{s}}\\ \check{B}^\dagger_{y\bar{s}.xcs}\end{pmatrix}=\begin{pmatrix}\hat{\beta}_{yx.cm}\\ \hat{B}_{ys.xc\bar{s}}\\ \hat{B}_{y\bar{s}.xcs}\end{pmatrix}+n\lambda_1\begin{pmatrix}-1 & \hat{B}_{\bar{s}x.sc} & \hat{B}_{\bar{z}x.zm}\\ \hat{B}_{xs.c\bar{s}} & \hat{B}_{\bar{s}s.xc} & \hat{B}_{\bar{z}s.xz\bar{s}}\\ \hat{B}_{x\bar{s}.cs} & -I_{q_{\bar{s}}} & \hat{B}_{\bar{z}\bar{s}.xsz}\end{pmatrix}$$
$$\times\begin{pmatrix}\zeta_1 s^{-1}_{xx.cm}\text{sign}(\check{\beta}^\dagger_{yx.cm})\\ \xi_1 S^{-1}_{\bar{s}\bar{s}.xcs}\gamma_{\bar{s}x.c}\odot\text{sign}(\check{B}^\dagger_{y\bar{s}.xcs})\\ (1-\zeta_1-\xi_1)S^{-1}_{\bar{z}\bar{z}.xmz}\gamma_{y\bar{z}.xmz}\odot\text{sign}(\check{B}^\dagger_{y\bar{z}.xmz})\end{pmatrix}. \qquad (A.16)$$

## B  Proof of Theorem 1

First, letting $\boldsymbol{v}$ be an active set from $\boldsymbol{x}\cup\boldsymbol{m}\cup\boldsymbol{c}$, for penalized estimators, such as ridge-type estimators and LASSO-type estimators, $\check{B}_{yv}$ and the ordinary least-squares estimator $\hat{B}_{yv}$ in the regression model of $Y$ on $\boldsymbol{V}$, from Zou (2006), we have

$$\check{B}_{yv}=(S_{vv}+\Gamma)^{-1}S_{vy}=(I_{vv}+S^{-1}_{vv}\Gamma)^{-1}\hat{B}_{yv}, \qquad (B.17)$$

$$\text{var}(\check{B}_{yv})=\sigma_{yy.v}(S_{vv}+\Gamma)^{-1}S_{vv}(S_{vv}+\Gamma)^{-1} \qquad (B.18)$$

approximately[1], where $\sigma_{yy.v}$, $S_{vv}$ and $\Gamma$ are the conditional variance of $Y$ given $\boldsymbol{V}$, the sum-of-products matrix of $\boldsymbol{v}$ and a semi-positive diagonal matrix determined by the penalty parameter in the regression model of $Y$ on $\boldsymbol{v}$, respectively.

Then, we prove Theorem 1 by the following steps:

**Step 1:** Compare $\text{var}(\check{B}_{yv})$ and $\text{var}(\hat{B}_{yv})$

**Step 2:** Compare $\text{var}(\check{B}^*_{mx.c}\hat{B}_{ym.c})$ and $\text{var}(\hat{B}_{mx.c}\hat{B}_{ym.c})$

**Step 3:** Compare $\text{var}(\check{B}^*_{mx.c}\hat{B}_{ym.c})$ and $\text{var}(\check{B}^*_{mx.c}\check{B}^*_{ym.c})$

---

[1] Given an active set, LASSO-type estimators can be replaced by ridge-type estimators through the local quadratic approximation (Zou, 2006).

**Step 4:** Compare $\mathrm{var}(\check{B}^*_{mx.c}\check{B}^*_{ym.c})$ and $\mathrm{var}(\check{\beta}^*_{yx.c})$

### Step 1: Comparison between $\mathrm{var}(\check{B}_{yv})$ and $\mathrm{var}(\hat{B}_{yv})$

We have

$$\begin{aligned}
\mathrm{var}(\check{B}_{yv}) - \mathrm{var}(\hat{B}_{yv}) &= \sigma_{yy.v}(S_{vv}+\Gamma)^{-1}S_{vv}(S_{vv}+\Gamma)^{-1} - \sigma_{yy.v}S_{vv}^{-1} \\
&= \sigma_{yy.v}(S_{vv}+\Gamma)^{-1}\left\{S_{vv} - (S_{vv}+\Gamma)S_{vv}^{-1}(S_{vv}+\Gamma)\right\}(S_{vv}+\Gamma)^{-1} \\
&= -\sigma_{yy.v}(S_{vv}+\Gamma)^{-1}\left\{2\Gamma + \Gamma S_{vv}^{-1}\Gamma\right\}(S_{vv}+\Gamma)^{-1}.
\end{aligned} \quad (\text{B.19})$$

Since $2\Gamma + \Gamma S_{vv}^{-1}\Gamma$ is a semi-positive definite matrix, if $\bm{v} = \bm{x} \cup \bm{c}$, then we derive

$$\bm{\omega}^T \mathrm{var}(\check{B}_{yv})\bm{\omega} \leq \bm{\omega}^T \mathrm{var}(\hat{B}_{yv})\bm{\omega}$$

for any $q_v$-dimensional nonzero vector $\bm{\omega}$, which leads to

$$\mathrm{var}(\check{\beta}_{yx.c}) \leq \mathrm{var}(\hat{\beta}_{yx.c}). \quad (\text{B.20})$$

Here,

$$\mathrm{var}(\hat{B}_{mx.c}\hat{B}_{ym.c}) \leq \mathrm{var}(\hat{\beta}_{yx.c}) \quad (\text{B.21})$$

was derived in Kuroki and Hayashi (2014, 2016).

### Step 2: Comparison between $\mathrm{var}(\check{B}^*_{mx.c}\hat{B}_{ym.c})$ and $\mathrm{var}(\hat{B}_{mx.c}\hat{B}_{ym.c})$

By applying the result of Step 1 to the relationship between $\check{B}^*_{mx.c}$ and $\hat{B}_{mx.c}$, we have

$$\begin{aligned}
&\mathrm{var}(\check{B}^*_{mx.c}\hat{B}_{ym.c}) - \mathrm{var}(\hat{B}_{mx.c}\hat{B}_{ym.c}) \\
&= \mathrm{var}(E(\check{B}^*_{mx.c}\hat{B}_{ym.c}|\bm{x},\bm{c},\bm{m})) + E(\mathrm{var}(\check{B}^*_{mx.c}\hat{B}_{ym.c}|\bm{x},\bm{c},\bm{m})) \\
&\quad - \mathrm{var}(E(\hat{B}_{mx.c}\hat{B}_{ym.c}|\bm{x},\bm{c},\bm{m})) - E(\mathrm{var}(\hat{B}_{mx.c}\hat{B}_{ym.c}|\bm{x},\bm{c},\bm{m})) \\
&= \mathrm{var}(\check{B}^*_{mx.c}B_{ym.c}) + \sigma_{yy.mc}E(\check{B}^*_{mx.c}S_{mm.c}^{-1}\check{B}^{*T}_{mx.c}) \\
&\quad - \mathrm{var}(\hat{B}_{mx.c}B_{ym.c}) - \sigma_{yy.mc}E(\hat{B}_{mx.c}S_{mm.c}^{-1}\hat{B}^T_{mx.c}) \\
&\leq \sigma_{yy.mc}\left\{E(\check{B}^*_{mx.c}S_{mm.c}^{-1}\check{B}^{*T}_{mx.c}) - E(\hat{B}_{mx.c}S_{mm.c}^{-1}\hat{B}^T_{mx.c})\right\}. \quad (\text{B.22})
\end{aligned}$$

Here, we have

$$\check{B}^*_{mx.c}S_{mm.c}^{-1}\check{B}^{*T}_{mx.c} - \hat{B}_{mx.c}S_{mm.c}^{-1}\hat{B}^T_{mx.c} = S_{xm.c}S_{mm.c}^{-1}S_{mx.c}\left((s_{xx.c}-\gamma)^{-2} - s_{xx.c}^{-2}\right) \leq 0.$$

Referring to equation (B.17), $\gamma$ is given by

$$\gamma = S_{xc}(S_{cc}+\Lambda_{cc})^{-1}S_{cx}, \quad (\text{B.23})$$

where $S_{cc}$, $S_{cx} = S_{xc}^T$ and $\Lambda_{cc}$ are the sum-of-products matrix of $\bm{c}$, the sum-of-cross-products matrix between $\bm{c}$ and $\bm{x}$, and the positive diagonal matrix determined by the penalty parameter in the regression model of $\bm{M}$ on $X$ and $\bm{C}$, respectively. This shows that

$$\mathrm{var}(\check{B}^*_{mx.c}\hat{B}_{ym.c}) \leq \mathrm{var}(\hat{B}_{mx.c}\hat{B}_{ym.c}).$$

### Step 3: Comparison between $\mathrm{var}(\check{B}^*_{mx.c}\hat{B}_{ym.c})$ and $\mathrm{var}(\check{B}^*_{mx.c}\check{B}^*_{ym.c})$

By applying the result of Step 1 to the relationship between $\check{B}^*_{ym.c}$ and $\hat{B}_{ym.c}$, we have

$$\begin{aligned}
&\mathrm{var}(\check{B}^*_{mx.c}\check{B}^*_{ym.c}) - \mathrm{var}(\check{B}^*_{mx.c}\hat{B}_{ym.c}) \\
&= \mathrm{var}(E(\check{B}^*_{mx.c}\check{B}^*_{ym.c}|\bm{x},\bm{c},\bm{m})) + E(\mathrm{var}(\check{B}^*_{mx.c}\check{B}^*_{ym.c}|\bm{x},\bm{c},\bm{m})) \\
&\quad - \mathrm{var}(E(\check{B}^*_{mx.c}\hat{B}_{ym.c}|\bm{x},\bm{c},\bm{m})) - E(\mathrm{var}(\check{B}^*_{mx.c}\hat{B}_{ym.c}|\bm{x},\bm{c},\bm{m})) \\
&\leq \mathrm{var}(E(\check{B}^*_{mx.c}\check{B}^*_{ym.c}|\bm{x},\bm{c},\bm{m})) - \mathrm{var}(E(\check{B}^*_{mx.c}\hat{B}_{ym.c}|\bm{x},\bm{c},\bm{m})) \\
&= B^T_{ym.c}\mathrm{var}(\check{B}^*_{mx.c}(I_{q_m,q_m} + S_{mm.c}^{-1}\Gamma)^{-1})B_{ym.c} - B^T_{ym.c}\mathrm{var}(\check{B}^*_{mx.c})B_{ym.c}. \quad (\text{B.24})
\end{aligned}$$

Here, $I_{q_m,q_m} + S_{mm.c}^{-1}\Gamma$ is a semipositive definite matrix and

$$\boldsymbol{\omega}^T(I_{q_m,q_m} + S_{mm.c}^{-1}\Gamma)\boldsymbol{\omega} - \boldsymbol{\omega}^T\boldsymbol{\omega} \geq 0 \tag{B.25}$$

holds for any $q_m$-dimensional nonzero vector $\boldsymbol{\omega}$, which leads to

$$\text{var}(\check{B}_{mx.c}^* \check{B}_{ym.c}^*) - \text{var}(\check{B}_{mx.c}^* \hat{B}_{ym.c}) \leq 0. \tag{B.26}$$

Thus, Steps 1~3 show that

$$\text{var}(\check{B}_{mx.c}^* \check{B}_{ym.c}^*) \leq \text{var}(\hat{B}_{mx.c} \hat{B}_{ym.c}) \leq \text{var}(\hat{\beta}_{yx.c})$$

holds approximately.

### Step 4: Comparison between $\text{var}(\check{B}_{mx.c}^* \check{B}_{ym.c}^*)$ and $\text{var}(\check{\beta}_{yx.c}^*)$

For the optimal semi-positive diagonal matrix $\Gamma$ to yield $\check{\beta}_{yx.c}^*$, which may not be optimal for $\check{B}_{mx.c}^* \hat{B}_{ym.c}$, we have

$$\sigma_{yy.xc} = \sigma_{yy.xmc} + B_{ym.xc}^T \Sigma_{mm.xc} B_{ym.xc} = \sigma_{yy.mc} + B_{ym.c}^T \Sigma_{mm.xc} B_{ym.c} \tag{B.27}$$

since $X$ is conditionally independent of $Y$ given $\boldsymbol{M} \cup \boldsymbol{C}$. Thus, from equation (B.23), we have

$$\text{var}(\check{\beta}_{yx.c}^*) - \text{var}(\check{B}_{mx.c}^* \hat{B}_{ym.c}) = \sigma_{yy.xc} E\left(\frac{s_{xx.c}}{(s_{xx.c} - \gamma)^2}\right)$$
$$- B_{ym.c}^T \text{var}(\check{B}_{mx.c}^*) B_{ym.c} - \sigma_{yy.mc} E(\check{B}_{mx.c}^* S_{mm.c}^{-1} \check{B}_{mx.c}^{*T})$$
$$\geq \sigma_{yy.mc} E\left(\frac{s_{xx.c} - S_{xm.c} S_{mm.c}^{-1} S_{mx.c}}{(s_{xx.c} - \gamma)^2}\right) = \sigma_{yy.mc} E\left(\frac{s_{xx.mc}}{(s_{xx.c} - \gamma)^2}\right) \geq 0. \tag{B.28}$$

Thus, together with the results of Step 3, we have

$$\text{var}(\check{\beta}_{yx.c}^*) \geq \text{var}(\check{B}_{mx.c}^* \hat{B}_{ym.c}) \geq \text{var}(\check{B}_{mx.c}^* \check{B}_{ym.c}^*) \tag{B.29}$$

approximately.

## C  Numerical Experiments

In this section, we conduct numerical experiments to compare the performances of LASSO, adaptive LASSO, Elastic Net, PAL$_1$MA, least squares methods, and PCM Selector.

### C.1  Loss Functions

#### Traditional Penalized Regression Analysis

For an $q_c$-dimensional regression vector $B_{yc.xm}$ and a $q_m$-dimensional regression vector $B_{ym.xc}$, let $B_y = (\beta_{yx.cm}, B_{yc.xm}^T, B_{ym.xc}^T)^T = (\beta_1, \beta_2, ..., \beta_{q_c+q_m+1})^T$ and $\lambda, \lambda' > 0$. First, the $L_1$-penalized loss function of adaptive LASSO (Zou, 2006) is defined as

$$\frac{1}{2n}\|\boldsymbol{y} - \boldsymbol{x}\beta_{yx.cm} - \boldsymbol{c}B_{yc.xm} - \boldsymbol{m}B_{ym.xc}\|_2^2 + \lambda\|\boldsymbol{\gamma} \odot B_y\|_1, \tag{C.30}$$

where $\boldsymbol{\gamma} = (\gamma_1, \gamma_2, ..., \gamma_{q_c+q_m+1})^T$ is a weight vector such that

$$\boldsymbol{\gamma} = \left(\frac{1}{|\tilde{\beta}_1|^\eta}, \frac{1}{|\tilde{\beta}_2|^\eta}, \ldots, \frac{1}{|\tilde{\beta}_{q_c+q_m+1}|^\eta}\right)^T \tag{C.31}$$

with tuning parameter $\eta \geq 0$, where $\tilde{\beta}_i$, $i = 1, 2, \ldots, q_c + q_m + 1$, is the standard ridge estimator of $B_y$ given a penalty parameter $\lambda'$ (Hoerl and Kennard, 1970). In particular, equation (C.30) is the $L_p$-penalized loss function of the standard LASSO (Tibshirani, 1996) when $\eta = 0$.

For $0 \leq \phi \leq 1$, the $L_p$-penalized loss function of the Elastic Net (Zou, 2006) is given by

$$\frac{1}{2n}\|\boldsymbol{y} - \boldsymbol{x}\beta_{yx.cm} - \boldsymbol{c}B_{yc.xm} - \boldsymbol{m}B_{ym.xc}\|_2^2 + \lambda\left((1-\phi)\|B_y\|_2^2 + \phi\|B_y\|_1^1\right). \quad (C.32)$$

**Partially Adaptive $L_p$-Regularization Multiple Regression Analysis (PAL$_p$MA)**

For an $n \times q_{\bar{z}}$ observation matrix $\bar{\boldsymbol{z}}$ and penalty parameter $\lambda_p \geq 0$, the $L_p$-penalized loss function of the original PAL$_p$MA (Nanmo and Kuroki, 2022) is given by

$$\frac{1}{2n}\|\boldsymbol{y} - \boldsymbol{x}\beta_{yx.c} - \boldsymbol{z}B_{yz.x\bar{z}} - \bar{\boldsymbol{z}}B_{y\bar{z}.xz}\|_2^2 + \lambda_p\|\boldsymbol{\gamma} \odot B_{y\bar{z}.xz}\|_p^p, \quad p = 1, 2 \quad (C.33)$$

where $\boldsymbol{\gamma} = (\gamma_1, \gamma_2, ..., \gamma_q)^T$ is a weight vector such that

$$\boldsymbol{\gamma} = \left(\frac{1}{|\tilde{\beta}_{y\bar{z}_1.xc}|^{\eta_p}}, \frac{1}{|\tilde{\beta}_{y\bar{z}_2.xc}|^{\eta_p}}, \ldots, \frac{1}{|\tilde{\beta}_{y\bar{z}_{q_{\bar{z}}}.xc}|^{\eta_p}}\right)^T \quad (C.34)$$

with tuning parameter $\eta_p \geq 0$, where $\tilde{B}_{y\bar{z}.xz}^T = (\tilde{\beta}_{y\bar{z}_1.xc}, \ldots, \tilde{\beta}_{y\bar{z}_{q_{\bar{z}}}.xc})^T$ is derived from

$$\begin{pmatrix} \tilde{\beta}_{yx.z\bar{z}} \\ \tilde{B}_{yz.x\bar{z}} \\ \tilde{B}_{y\bar{z}.xz} \end{pmatrix} = \begin{pmatrix} s_{xx} & S_{xz} & S_{x\bar{z}} \\ S_{zx} & S_{zz} & S_{z\bar{z}} \\ S_{\bar{z}x} & S_{\bar{z}z} & n\lambda I_{q_{\bar{z}}} + S_{\bar{z}\bar{z}} \end{pmatrix}^{-1} \begin{pmatrix} s_{xy} \\ S_{zy} \\ S_{\bar{z}y} \end{pmatrix} \quad (C.35)$$

given a penalty parameter $\lambda > 0$. Letting $\check{B}_y^{\dagger T} = (\check{\beta}_{yx.c}^{\dagger}, \check{B}_{yz.x\bar{z}}^{\dagger T}, \check{B}_{y\bar{z}.xz}^{\dagger T})^T$ be the estimator that minimizes the $L_p$-penalized loss function (C.33) for $p = 1$, the estimators of the total effect of PAL$_1$MA are defined by correcting the bias term of $\check{\beta}_{yx.c}^{\dagger}$ using $\boldsymbol{\gamma}$. Here, the stability of the estimated $\boldsymbol{\gamma}$ may have an effect on the bias correction. To avoid this difficulty, in the numerical experiments, we apply the following standardized weight vector to $\boldsymbol{\gamma}$ in equation (C.33):

$$\boldsymbol{\gamma}' = \left(\sum_{i=1}^{q_{\bar{z}}} \frac{1}{|\tilde{\beta}_{y\bar{z}_i.xc}|^{\eta_p}}\right)^{-1} \left(\frac{1}{|\tilde{\beta}_{y\bar{z}_1.xc}|^{\eta_p}}, \frac{1}{|\tilde{\beta}_{y\bar{z}_2.xc}|^{\eta_p}}, \ldots, \frac{1}{|\tilde{\beta}_{y\bar{z}_{q_{\bar{z}}}.xc}|^{\eta_p}}\right)^T. \quad (C.36)$$

Then, in the framework of PAL$_1$MA, the total effect is estimated by

$$\check{\beta}_{yx.c}^* = \check{\beta}_{yx.c}^{\dagger} - \frac{n\lambda_1}{\tilde{s}_{xx.c}^{\dagger}} \tilde{B}_{x\bar{z}.z}^{\dagger T} \boldsymbol{\gamma} \odot \text{sign}(\check{B}_{y\bar{z}.xz}^{\dagger}), \quad (C.37)$$

$$\tilde{s}_{xx.c}^{\dagger} = \|\boldsymbol{x} - \boldsymbol{z}\tilde{B}_{xz.\bar{z}}^{\dagger} - \bar{\boldsymbol{z}}\tilde{B}_{x\bar{z}.z}^{\dagger}\|_2^2 \quad (C.38)$$

where $\tilde{B}_{x\bar{z}.z}^{\dagger}$ and $\tilde{B}_{xz.\bar{z}}^{\dagger}$ are PAL$_2$MA estimators derived from the $L_p$-penalized loss function with a standardized weight vector $\boldsymbol{\gamma}''$, a penalty parameter $\lambda' \geq 0$ and a tuning parameter $\eta_2 > 0$ such that $\boldsymbol{x}$ and $\boldsymbol{y}$ are replaced by an empty set and $\boldsymbol{x}$, respectively, in equation (C.33) for $p = 2$.

Note that the R package "`glmnet`" (version 4.1.8) (Friedman et al., 2023) is utilized to perform LASSO, adaptive LASSO, Elastic Net, PAL$_1$MA and PCM Selector. All experiments were carried out on an Intel Core i7-1360P CPU running at 2.20 GHz.

### C.2 Parameter settings

For simplicity, letting $X$ and $Y$ be the treatment variable and the response variable, respectively, consider linear SCMs with 18 explanatory variables for $Y$ in the form

$$\left.\begin{aligned} Y &= \alpha_{ys}S + \alpha_{yz}Z + \overline{\boldsymbol{S}}A_{y\bar{s}} + \overline{\boldsymbol{Z}}A_{y\bar{z}} + \epsilon_y \\ \overline{\boldsymbol{S}} &= XA_{\bar{s}x} + SA_{\bar{s}s} + ZA_{\bar{s}z} + \boldsymbol{\epsilon}_{\bar{s}} \\ S &= \alpha_{sx}X + \alpha_{sz}Z + \epsilon_s \\ X &= \alpha_{xz}Z + \epsilon_x \end{aligned}\right\} \quad (C.39)$$

where $\overline{\boldsymbol{S}}$ and $\overline{\boldsymbol{Z}}$ include 5 and 10 variables, respectively. In addition, $A_{y\bar{s}} = (\alpha_{y\bar{s}_1}, \alpha_{y\bar{s}_2}, \ldots, \alpha_{y\bar{s}_5})^T$, $A_{y\bar{z}} = (\alpha_{y\bar{z}_1}, \alpha_{y\bar{z}_2}, \ldots, \alpha_{y\bar{z}_{10}})^T$, $A_{\bar{s}x} = (\alpha_{\bar{s}_1x}, \alpha_{\bar{s}_2x}, \ldots, \alpha_{\bar{s}_5x})$, $A_{\bar{s}s} =$

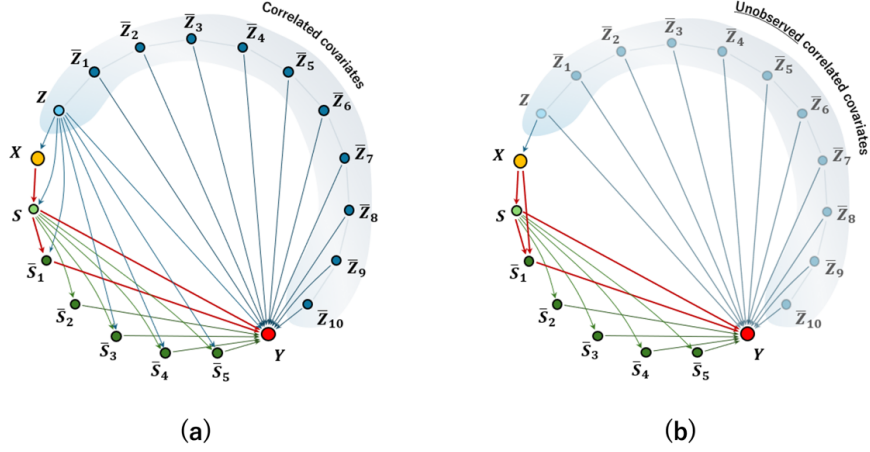

(a)  (b)

Fig. A. Causal diagram. The red arrows show the total effect of interest. $X$: treatment variable; $Y$: response variable; $S$: intermediate variable that can be selected using prior causal knowledge; $\overline{S} = \{\overline{S}_1, \overline{S}_2, \ldots, \overline{S}_5\}$: a set of intermediate variables for which it is uncertain which element should be added to evaluate the total effects; $Z$: covariate that can be selected using prior causal knowledge; $\overline{Z} = \{\overline{Z}_1, \overline{Z}_2, \ldots, \overline{Z}_{10}\}$: a set of covariates for which it is uncertain which element should be added to evaluate the total effects.

$(\alpha_{\overline{s}_1 s}, \alpha_{\overline{s}_2 s}, \ldots, \alpha_{\overline{s}_5 s})$ and $A_{\overline{s} z} = (\alpha_{\overline{s}_1 z}, \alpha_{\overline{s}_2 z}, \ldots, \alpha_{\overline{s}_5 z})$. Fig. A (a) shows that (i) $S$ satisfies the front-door-like criterion relative to $(X, Y)$ with $Z$ and (ii) $Z$ satisfies the back-door criterion relative to $(X, Y)$. Fig. A (b) shows that (i) $\{S, \overline{S}_1\}$ satisfies the front-door criterion relative to $(X, Y)$ and (ii) $C = \{Z, \overline{Z}\}$ satisfies the back-door criterion relative to $(X, Y)$ but is unobserved. Here, $S$ and $\{S, \overline{S}_1\}$ are the minimally sufficient sets of intermediate variables that satisfies the front-door-like criterion for Fig. A (a), and satisfies the front-door criterion for Fig. A (b), respectively.

To construct the population variance-covariance matrix with the linear SCMs (C.39), we first assigned one of 0.1 and 0.8 to $\alpha_{xz}$ and $\alpha_{sx}$ depending on Fig. A (a) and to $\alpha_{sx}$ depending on Fig. A (b). Here, multicollinearity may occur between $X$ and the covariates satisfying the back-door criterion or intermediate variables satisfying the front-door-like criterion when we assign 0.8 to the direct effects on $X$ but may not occur when we assign 0.1 to the direct effects on $X$. The direct effect $\alpha_{ys}$ was set to 0.4, the direct effects $\alpha_{\overline{s}_1 s}, \alpha_{\overline{s}_2 s}, \ldots, \alpha_{\overline{s}_5 s}(\in A_{\overline{s} s})$ were all set to 0.2, and the direct effects $\alpha_{\overline{s}_2 x}, \alpha_{\overline{s}_3 x}, \ldots, \alpha_{\overline{s}_5 x}(\in A_{\overline{s} x})$ were all set to 0 in the both settings Fig. A (a) and (b). The direct effects $\alpha_{\overline{s}_1 z}, \alpha_{\overline{s}_2 z}, \ldots, \alpha_{\overline{s}_5 z}(\in A_{\overline{s} z})$ were all set to 0.2 in the settings Fig. A (a) and were all set to 0 in the settings Fig. A (b). In addition, the direct effects $\alpha_{y\overline{z}_1}, \alpha_{y\overline{z}_2}, \ldots, \alpha_{y\overline{z}_{10}}(\in A_{y\overline{z}})$ and $\alpha_{y\overline{s}_2}, \alpha_{y\overline{s}_3}, \alpha_{y\overline{s}_4}, \alpha_{y\overline{s}_5}(\in A_{y\overline{s}})$ were randomly and independently generated according to a uniform distribution on the interval $[-0.2, 0.2]$. The other direct effects are given in Table A. In addition, the population variance-covariance

Table A. Direct effects

(a) $S$ satisfies the front-door-like criterion relative to $(X, Y)$ with $Z$
    $Z$ satisfies the back-door criterion relative to $(X, Y)$

| Fig. A (a) | $\alpha_{xz}$ | $\alpha_{sx}$ | $\alpha_{\overline{s}_1 x}$ | $\alpha_{yz}$ | $\alpha_{sz}$ | $\alpha_{y\overline{s}_1}$ |
|---|---|---|---|---|---|---|
| $(a_1)$ | 0.1 | 0.1 | 0.0 | 0.2 | 0.2 | $U([-0.2, 0.2])$ |
| $(a_2)$ | 0.1 | 0.8 | 0.0 | 0.2 | 0.2 | $U([-0.2, 0.2])$ |
| $(a_3)$ | 0.8 | 0.1 | 0.0 | 0.2 | 0.2 | $U([-0.2, 0.2])$ |
| $(a_4)$ | 0.8 | 0.8 | 0.0 | 0.2 | 0.2 | $U([-0.2, 0.2])$ |

(b) $\{S, \overline{S}_1\}$ satisfies the front-door criterion relative to $(X, Y)$

| Fig. A (b) | $\alpha_{xz}$ | $\alpha_{sx}$ | $\alpha_{\overline{s}_1 x}$ | $\alpha_{yz}$ | $\alpha_{sz}$ | $\alpha_{y\overline{s}_1}$ |
|---|---|---|---|---|---|---|
| $(b_1)$ | 0.2 | 0.1 | 0.2 | 0.0 | 0.0 | 0.2 |
| $(b_2)$ | 0.2 | 0.8 | 0.2 | 0.0 | 0.0 | 0.2 |

$U([-0.2, 0.2])$: direct effects determined by random numbers from the uniform distribution on the interval $[-0.2, 0.2]$.

matrices of the covariates $\boldsymbol{C}$ were randomly generated using the "`randcorr`" package (available from `https://www.rdocumentation.org/packages/randcorr/versions/1.0/topics/randcor r-package`) according to Pourahmadi and Wang (2015). In addition, we assume that (i) the random disturbances $\epsilon_x$, $\epsilon_s$, $\boldsymbol{\epsilon_{\overline{s}}}$ and $\epsilon_y$ independently follow normal distributions with mean 0 and variance or variance-covariance matrix $\sigma_{xx.c}$, $\sigma_{ss.cz}$, $\Sigma_{\overline{ss}.xsz}$, and $\sigma_{yy.cm}$, respectively, and (ii) the random disturbances are independent of their non-descendants. Here, the variances and variance-covariance matrices $\sigma_{xx.c}$, $\sigma_{ss.cz}$, $\Sigma_{\overline{ss}.xsz}$, and $\sigma_{yy.cm}$ are determined to satisfy the criterion that the variance of each variable in the corresponding linear SCM equals one.

Regarding the tuning penalty parameters for LASSO, adaptive LASSO, Elastic Net, PAL$_1$MA, and PCM Selector, the "`glmnet`" package is utilized in this paper. For tuning the penalty parameter in the $L_p$-penalized loss function, we referred to the search range $(0, \sqrt{\log(q')/n}]$ proposed by Bühlmann and van de Geer (2011), where $q'$ is the number of variables corresponding to the penalized part of regression coefficients. In addition, the search ranges of the other tuning parameters were set to $\zeta_p, \xi_p \in \{0.01, 0.02, \ldots, 0.99\}$ for PCM Selector ($\zeta_p + \xi_p \in [0, 1]$ for $p = 1, 2$) and $\eta, \eta_1, \eta_2 \in \{0.1, 0.2, 0.3, ..., 2.9, 3.0\}$ for adaptive LASSO and PAL$_p$MA. The mixing parameter $\phi$ of Elastic Net was set to $\phi \in \{0.01, 0.02, 0.03, ..., 0.98, 0.99\}$.

Based on the abovementioned parameter ranges, we conducted all possible selections based on threefold cross-validation to determine the combination of parameters based on the mean squared error. The results of the parameter tuning are shown in Table B. Note that the parameter settings of PAL$_1$MA and PCM Selector in this paper are somewhat empirical; i.e., they may not be determined as optimally as in other penalized regression analyses. The development of optimal parameter tuning for PAL$_1$MA and PCM Selector is left for future work.

Table B. Parameter settings based on threefold cross-validation

| Fig. A (a) | LASSO | adaptive LASSO | | | Elastic Net | | | PAL$_1$MA | | | | PCM Selector | | | | | |
|---|---|---|---|---|---|---|---|---|---|---|---|---|---|---|---|---|---|
| | $\lambda$ | $\lambda$ | $\eta$ | $\lambda'$ | $\lambda$ | $\phi$ | | $\lambda_1$ | $\eta_1$ | $\lambda$ | | $\zeta_1$ | $\lambda_1$ | $\xi_1$ | $\rho_1$ | $\lambda$ | $\rho$ | $\tau_{yx}$ |
| ($a_1$) | 0.407 | 0.269 | 0.100 | 451.940 | 0.407 | 0.910 | 0.392 | 1.300 | 249.858 | 0.400 | 0.009 | 0.030 | 0.213 | 3.377 | 52.653 | 0.045 |
| ($a_2$) | 0.142 | 0.151 | 0.100 | 5.242 | 0.142 | 0.920 | 0.306 | 1.200 | 4.516 | 0.610 | 0.013 | 0.010 | 0.213 | 3.107 | 64.798 | 0.362 |
| ($a_3$) | 0.407 | 0.407 | 0.100 | 395.298 | 0.399 | 0.900 | 0.294 | 1.200 | 308.926 | 0.390 | 0.017 | 0.050 | 0.213 | 3.157 | 69.484 | 0.045 |
| ($a_4$) | 0.028 | 0.073 | 1.900 | 4.561 | 0.033 | 0.930 | 0.020 | 0.900 | 2.070 | 0.270 | 0.013 | 0.190 | 0.213 | 3.565 | 44.102 | 0.362 |

| Fig. A (b) | PCM Selector | | | | | | | |
|---|---|---|---|---|---|---|---|---|
| | $\lambda_1$ | $\zeta_1$ | $\xi_1$ | $\rho_1$ | $\lambda$ | $\rho$ | $\tau_{yx}$ |
| ($b_1$) | 0.076 | 0.000 | 1.000 | - | 253.515 | - | 0.085 |
| ($b_2$) | 0.346 | 0.000 | 1.000 | - | 3.726 | - | 0.402 |

$\rho$, $\rho_1$, $\lambda$, $\lambda_1$, $\lambda'$: penalty parameters; $\eta$, $\eta_1$, $\zeta_1$, $\xi_1$, $\xi_2$: tuning parameters; $\phi$: mixing parameter; $\tau_{yx}$: total effect of $X$ on $Y$.

Table C. Parameter settings in replications

| Fig. A (a) | adaptive LASSO | PAL$_1$MA | | | PCM Selector | | | | | |
|---|---|---|---|---|---|---|---|---|---|---|
| | $\lambda'$ | $\lambda$ | $\lambda_2$ | $\eta_2$ | $\lambda$ | $\rho$ | $\lambda_2$ | $\xi_2$ | $\rho_2$ | $\rho'_2$ |
| ($a_1$) | 167.620 | 131.372 | 0.000 | 0.000 | 128.569 | 34.981 | 0.000 | 0.000 | 0.000 | 0.000 |
| ($a_2$) | 127.820 | 114.470 | 0.000 | 0.000 | 126.195 | 33.734 | 0.000 | 0.000 | 0.000 | 0.000 |
| ($a_3$) | 150.285 | 130.230 | 0.000 | 0.000 | 124.013 | 37.411 | 0.000 | 0.000 | 0.000 | 0.000 |
| ($a_4$) | 54.846 | 109.683 | 0.000 | 0.000 | 116.405 | 35.220 | 0.000 | 0.030 | 0.000 | 0.000 |

| Fig. A (b) | PAL$_1$MA | | PCM Selector | | | | |
|---|---|---|---|---|---|---|---|
| | $\lambda$ | $\lambda_2$ | $\rho$ | $\lambda_2$ | $\xi_2$ | $\rho_2$ | $\rho'_2$ |
| ($b_1$) | 69.487 | 0.000 | - | 0.000 | 1.000 | 0.000 | - |
| ($b_2$) | 70.590 | 0.000 | - | 0.000 | 1.000 | 0.000 | - |

$\rho$, $\rho_2$, $\rho'_2$, $\lambda$, $\lambda'$, $\lambda_2$: penalty parameters; $\eta_2$, $\xi_2$: tuning parameters. All parameter values in this table are shown as means in 5000 replications.

9

### C.3 Analysis

For 5000 replications, we generated 15 random samples of 18 variables from a multivariate normal distribution with a zero mean vector and the population variance-covariance matrix generated by the above procedure. Tables D and E show the numerical results by LASSO, adaptive LASSO, Elastic Net, PAL$_1$MA, the OLS method, the TSLS method, and PCM Selector based on Table A. Here, the TSLS methods are based on front-door-like criterion in Setting (a) and based on front-door criterion in Setting (b). In addition, for the OLS and TSLS methods, we select a set of covariates $C$ in Fig. A (a). In Fig. A (b), it is assumed that a set of covariates is not observed, and thus the total effect can not be estimated by using the back-door criterion.

From Figs. B and C and Tables D and E, we make the following observations:

1. When the total effect is close to zero, the coincidence rates between the signs of the estimated total effects and the true total effects are low for LASSO, adaptive LASSO, and Elastic Net. This would be serious because it provides such a misleading interpretation that the external intervention of the treatment variable $X$ does not have no effect on the change of $Y$. In contrast, the coincidence rates of PAL$_1$MA and PCM Selector are still higher than those of the other regression analyses. Here, when the true total effect is far from zero, the coincidence rates are high for each regression analysis.

2. The OLS method provides an unbiased estimator of the total effect through a whole set of covariates that satisfy the back-door criterion, and the TSLS method also provides an unbiased estimator of the total effect through a minimally sufficient set of intermediate variables that satisfy the front-door-like criterion with a whole set of covariates. Given this finding, the estimators from the penalized regression analysis are expected to be close to both the OLS estimate and the TSLS estimate. However, from Tables D and E, the estimates from PAL$_1$MA are close to the OLS estimates, but the estimates from PCM Selector are close to the TSLS estimates (not including $x$). The difference between the OLS (PAL$_1$MA) estimate and the TSLS (PCM Selector) estimate may be due to the small sample size problem. Here, note that both PCM Selector and PAL$_1$MA provide better estimation accuracy than the OLS and the TSLS methods in most cases.

3. The variances of the estimated total effects from PCM Selector are larger than those from the other traditional penalized regression analyses but smaller than those from OLS and TSLS methods in most cases. In addition, it seems that PAL$_1$MA provides better estimation accuracy than PCM Selector in some cases. This seems to contradict Theorem 1, but it is not, because Theorem 1 is derived under the assumption that PAL$_1$MA and PCM Selector utilize the same set of covariates and the same weight matrix.

4. PCM Selector provides consistent or less biased estimators of the total effect than other regression analyses.

Overall, the coincidence rates between the signs of the estimated total effects and the true total effect from PCM Selector seem equal to or higher than those from the other regression analyses. In addition, in some cases of Fig. A, PCM Selector may not select a set of covariates/intermediate variables that satisfies the front-door-like criterion, and such a missing covariate/intermediate variable may provide biased estimates of the total effects. Then, PCM Selector may reverse the direction of the regression coefficient in such situations. However, regarding PCM Selector, such a drawback is eliminated by selecting smaller values of the penalized parameters based on the whole set of covariates and intermediate variables. That is to say, since a set of covariates and intermediate variables is selected based on the sign of the estimated total effect of $X$ on $Y$ by PCM Selector with the smaller penalized parameter values, we can verify that the lack of sufficient confounders and intermediate variables does not interfere with the qualitative interpretation of the total effects. Thus, PCM Selector and PAL$_1$MA can provide less biased estimators than the other penalized regression analyses in most cases. This indicates that the estimation of the total effect

by PCM Selector does not lead to a misleading qualitative interpretation compared to the standard penalized regression analysis.

Finally, we would like to emphasize that most of the current penalized regression analyses, such as LASSO, adaptive LASSO, Elastic Net, and $PAL_pMA$, can not be applied to evaluate the total effects when a set of covariates satisfying the back-door criterion cannot be observed. In contrast, although we discussed the performances of LASSO, adaptive LASSO, Elastic Net, $PAL_1MA$ and PCM Selector separately, PCM Selector provides a wider class, including LASSO, adaptive LASSO, and $PAL_pMA$. In addition, PCM Selector is also applicable to some situations where a set of covariates satisfying the back-door criterion cannot be observed.

Table D. Results based on cross-validation.

(a) $S$ satisfies the front-door-like criterion relative to $(X, Y)$ with $Z$
$Z$ satisfies the back-door criterion relative to $(X, Y)$

($a_1$) $\tau_{yx} = 0.045$

|  | Mean | SD | Bias | RMSE | Sign |
|---|---|---|---|---|---|
| LASSO | 0.005 | 0.034 | -0.040 | 0.053 | 0.068 |
| adaptive LASSO | 0.018 | 0.087 | -0.028 | 0.091 | 0.192 |
| Elastic Net | 0.006 | 0.041 | -0.039 | 0.056 | 0.093 |
| PAL$_1$MA | 0.058 | 0.306 | 0.013 | 0.306 | 0.578 |
| PCM Selector | 0.047 | 0.338 | 0.001 | 0.338 | 0.566 |
| Front-door-like (including $x$) | 0.029 | 0.597 | -0.016 | 0.598 | 0.543 |
| Front-door-like (not including $x$) | 0.042 | 0.418 | -0.003 | 0.418 | 0.559 |
| Back-door | 0.052 | 0.633 | 0.007 | 0.633 | 0.551 |

($a_2$) $\tau_{yx} = 0.362$

|  | Mean | SD | Bias | RMSE | Sign |
|---|---|---|---|---|---|
| LASSO | 0.250 | 0.191 | -0.112 | 0.221 | 0.821 |
| adaptive LASSO | 0.256 | 0.196 | -0.106 | 0.223 | 0.817 |
| Elastic Net | 0.256 | 0.191 | -0.106 | 0.219 | 0.833 |
| PAL$_1$MA | 0.417 | 0.263 | 0.055 | 0.269 | 0.933 |
| PCM Selector | 0.409 | 0.500 | 0.047 | 0.503 | 0.833 |
| Front-door-like (including $x$) | 0.425 | 2.258 | 0.063 | 2.258 | 0.614 |
| Front-door-like (not including $x$) | 0.419 | 0.479 | 0.057 | 0.483 | 0.834 |
| Back-door | 0.419 | 0.547 | 0.057 | 0.550 | 0.817 |

($a_3$) $\tau_{yx} = 0.045$

|  | Mean | SD | Bias | RMSE | Sign |
|---|---|---|---|---|---|
| LASSO | 0.013 | 0.045 | -0.033 | 0.056 | 0.117 |
| adaptive LASSO | 0.017 | 0.057 | -0.028 | 0.063 | 0.138 |
| Elastic Net | 0.017 | 0.054 | -0.028 | 0.060 | 0.156 |
| PAL$_1$MA | 0.054 | 0.792 | 0.009 | 0.792 | 0.528 |
| PCM Selector | 0.036 | 0.718 | -0.010 | 0.718 | 0.526 |
| Front-door-like (including $x$) | -0.008 | 1.577 | -0.053 | 1.578 | 0.515 |
| Front-door-like (not including $x$) | 0.030 | 1.051 | -0.015 | 1.051 | 0.524 |
| Back-door | 0.054 | 1.591 | 0.009 | 1.591 | 0.532 |

($a_4$) $\tau_{yx} = 0.362$

|  | Mean | SD | Bias | RMSE | Sign |
|---|---|---|---|---|---|
| LASSO | 0.306 | 0.350 | -0.056 | 0.355 | 0.681 |
| adaptive LASSO | 0.351 | 0.719 | -0.011 | 0.719 | 0.683 |
| Elastic Net | 0.301 | 0.331 | -0.061 | 0.337 | 0.689 |
| PAL$_1$MA | 0.375 | 0.796 | 0.013 | 0.796 | 0.703 |
| PCM Selector | 0.370 | 0.952 | 0.008 | 0.952 | 0.677 |
| Front-door-like (including $x$) | 0.390 | 8.281 | 0.027 | 8.281 | 0.535 |
| Front-door-like (not including $x$) | 0.380 | 1.187 | 0.018 | 1.187 | 0.660 |
| Back-door | 0.380 | 1.312 | 0.018 | 1.312 | 0.658 |

Mean: sample mean; SD: standard deviation; Bias: bias between the true value and the sample mean; RMSE: root mean squared error: Sign: coincidence rate between the signs of the true value and the estimates; Front-door-like (including $x$): the treatment variable $X$, the intermediate variable $S$ and the set of covariates $\boldsymbol{C}$ are used for the front-door-like criterion; Front-door-like (not including $x$): intermediate variable $S$ and the set of covariates $\boldsymbol{C}$ are used for the front-door-like criterion; Back-door: the set of covariates $\boldsymbol{C}$ are used for the back-door criterion. $\tau_{yx}$ shows true value of total effect.

Table E. Results based on cross-validation.

(b) $\{S, \overline{S}_1\}$ satisfies the front-door criterion relative to $(X, Y)$
$\boldsymbol{C}$ satisfies the back-door criterion relative to $(X, Y)$

| | ($b_1$) $\tau_{yx} = 0.085$ | | | | |
|---|---|---|---|---|---|
| | Mean | SD | Bias | RMSE | Sign |
| PCM Selector | 0.083 | 0.204 | -0.002 | 0.204 | 0.666 |
| Front-door (minimal) | 0.091 | 0.172 | 0.006 | 0.173 | 0.713 |
| Front-door (whole) | 0.090 | 0.244 | 0.004 | 0.244 | 0.654 |

| | ($b_2$) $\tau_{yx} = 0.402$ | | | | |
|---|---|---|---|---|---|
| | Mean | SD | Bias | RMSE | Sign |
| PCM Selector | 0.448 | 0.549 | 0.046 | 0.551 | 0.808 |
| Front-door (minimal) | 0.468 | 0.552 | 0.066 | 0.556 | 0.818 |
| Front-door (whole) | 0.462 | 0.692 | 0.060 | 0.694 | 0.770 |

Mean: sample mean; SD: standard deviation; bias: bias between the true value and the sample mean; RMSE: root mean squared error: Sign: coincidence rate between the signs of the true value and the estimates; Front-door (minimal): the minimal subset of intermediate variables $\{S, \overline{S}_1\}$ is used for the front-door criterion. Front-door (whole): the set of intermediate variables $\boldsymbol{M}$ is used for the front-door criterion. $\tau_{yx}$ shows the true value of the total effect.

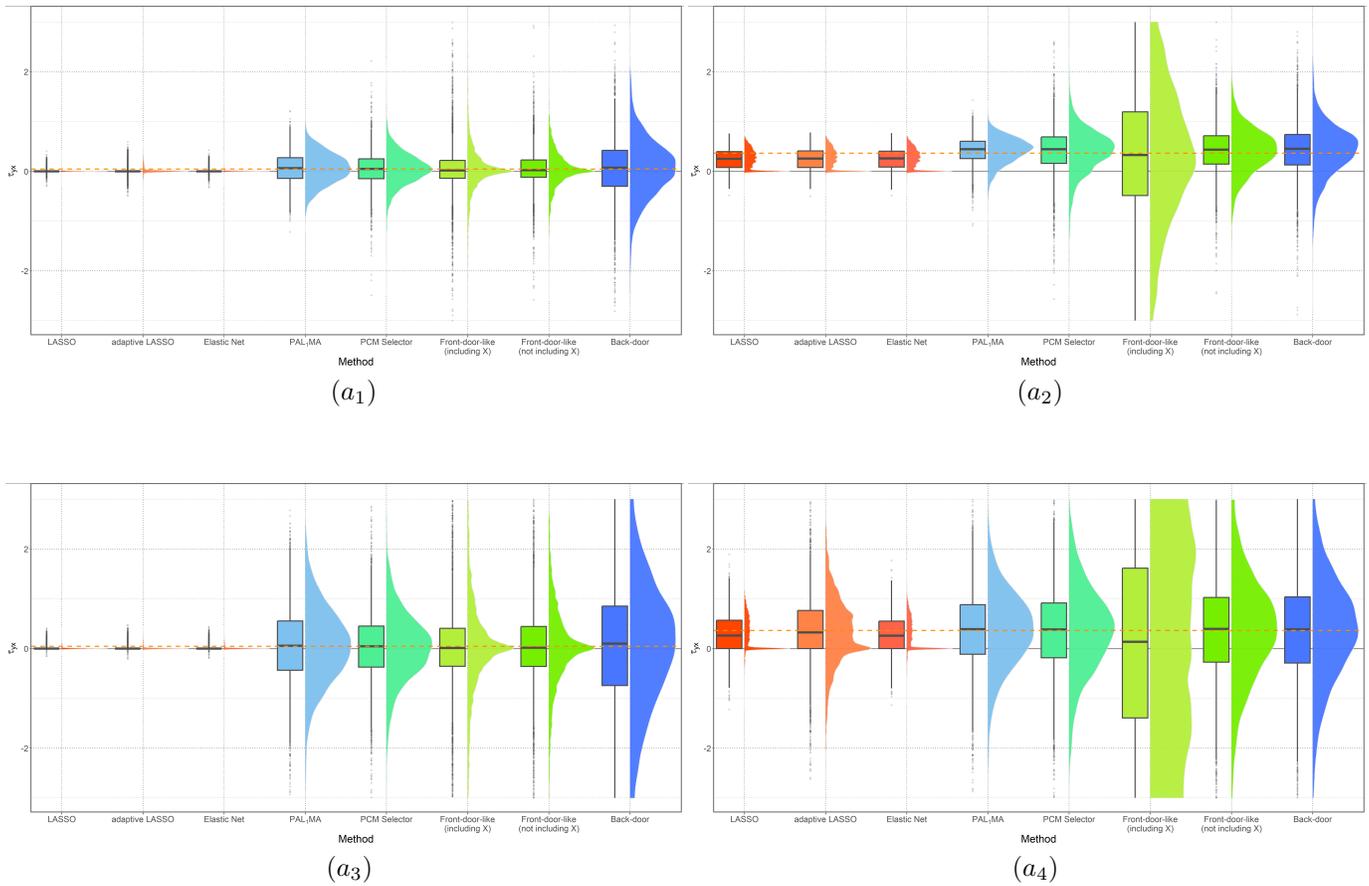

(a) $S$ satisfies the front-door-like criterion relative to $(X, Y)$ with $Z$
$Z$ satisfies the back-door criterion relative to $(X, Y)$

Fig. B. Violin plots of the estimated total effects based on 5000 replications from the numerical experiments. The dashed lines show the true total effects.

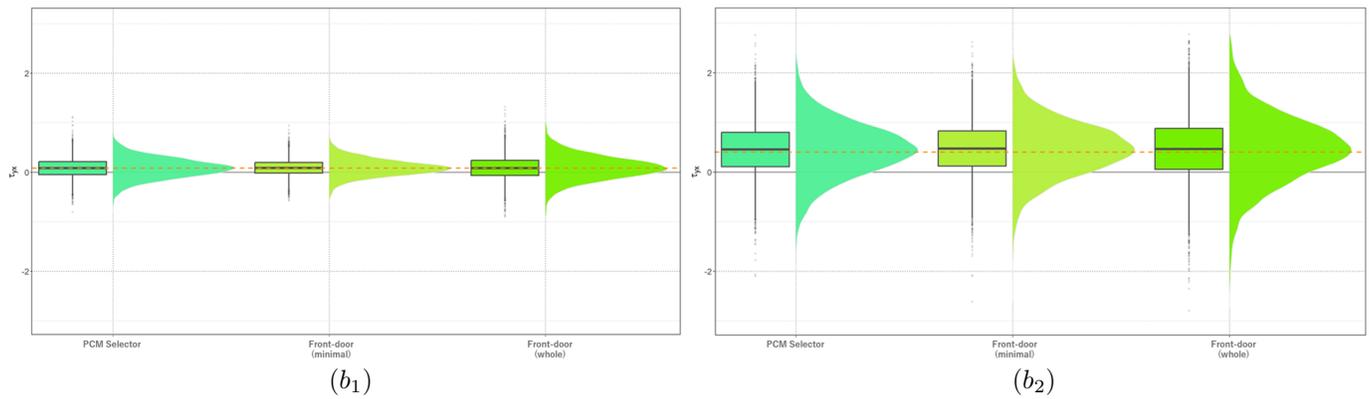

(b) $\{S, \overline{S}_1\}$ satisfies the front-door criterion relative to $(X, Y)$
$C$ satisfies the back-door criterion relative to $(X, Y)$

Fig. C. Violin plots of the estimated total effects based on 5000 replications from the numerical experiments. The dashed lines show the true total effects.

# D  Application to a Real-World Dataset

## D.1  Problem Setting

In this section, we apply LASSO, adaptive LASSO, Elastic Net, PAL$_1$MA, PCM Selector, the OLS method, and the TSLS method to a case study of setting up coating conditions for car bodies, as reported by Okuno et al. (1986) and reanalyzed by Kuroki (2012) and Nanmo and Kuroki (2021).

According to Kuroki (2012), car bodies are coated to improve both the rust protection quality and the visual appearance. A certain coating thickness must be ensured in the coating process. At the time of the study, this process was conducted by operators who sprayed the car bodies with paint, which depended on the operators' skills and could lead to low transfer efficiency. Okuno et al. (1986) collected nonexperimental data on the coating process to examine the process conditions and to increase the transfer efficiency. The sample size was 38, and the dataset is available from Okuno et al. (1986). In addition, the observed variables of interest are as follows:

<p align="center">Process conditions</p>

The dilution ratio ($X_1$), degree of viscosity ($X_2$), gun speed ($X_3$), spray distance ($X_4$), air pressure ($X_5$), pattern width ($X_6$), fluid output ($X_7$), paint temperature ($X_8$), temperature ($X_9$), and degree of moisture ($X_{10}$)

<p align="center">Response</p>

<p align="center">The transfer efficiency ($Y$).</p>

Table F shows the randomly selected data from the whole dataset given by Okuno et al. (1986). Note that our discussion is based on Table F and considers a situation where the OLS method with the all-variable selection procedure cannot be applied.

According to Kuroki (2012), there are some differences among these variables in terms of the controllability level: $X_1, X_2, ..., X_6$ can be controlled (i.e., treatment variables); $X_7$ and $X_8$ result from other factors and are difficult to control; and $X_9$ and $X_{10}$ are environmental conditions that cannot be controlled. Here, we assume that the cause-effect relationships in the coating process are as shown in Fig. D. From Fig. D, sets of covariates, including $X_{10}$, satisfy the back-door criterion relative to $(X_1, Y)$. In addition, $X_2$, $X_7$ and $\{X_2, X_7\}$ satisfy the front-door-like criterion relative to $(X_1, Y)$. For details on this case study, refer to Kuroki (2012).

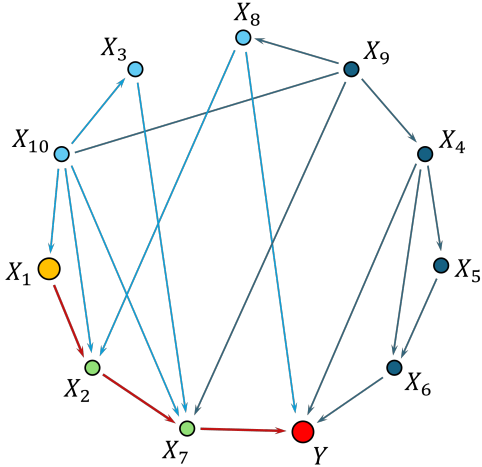

Fig. D. Graphical representation of the case study of setting up coating conditions for car bodies. The red-directed path shows the total effect of interest.

Table F. Randomly selected data from Okuno et al. (1986).

| No. | $X_1$ | $X_2$ | $X_3$ | $X_4$ | $X_5$ | $X_6$ | $X_7$ | $X_8$ | $X_9$ | $X_{10}$ | $Y$ |
|---|---|---|---|---|---|---|---|---|---|---|---|
| 1 | 16.7 | 35.0 | 4.9 | 40.0 | 5.0 | 3.9 | 168.0 | 25.0 | 20.0 | 25.0 | 28.7 |
| 2 | 16.7 | 28.0 | 8.3 | 40.0 | 2.8 | 5.0 | 112.0 | 32.0 | 22.0 | 29.0 | 19.6 |
| 3 | 33.0 | 25.5 | 6.5 | 40.0 | 4.0 | 4.0 | 276.0 | 20.0 | 22.5 | 25.0 | 17.8 |
| 4 | 44.0 | 29.5 | 6.5 | 30.0 | 2.1 | 5.0 | 120.0 | 6.7 | 7.0 | 30.0 | 21.7 |
| 5 | 33.0 | 28.3 | 8.3 | 40.0 | 2.0 | 3.0 | 318.0 | 20.0 | 19.0 | 30.0 | 22.8 |
| 6 | 44.0 | 29.5 | 6.5 | 30.0 | 4.9 | 5.0 | 180.0 | 6.7 | 7.0 | 30.0 | 54.8 |
| 7 | 16.7 | 28.0 | 8.3 | 40.0 | 4.5 | 1.0 | 128.0 | 33.0 | 10.5 | 39.0 | 19.5 |
| 8 | 44.0 | 24.2 | 5.0 | 30.0 | 2.0 | 5.0 | 108.0 | 28.0 | 22.5 | 25.0 | 19.8 |
| 9 | 16.7 | 50.0 | 5.0 | 40.0 | 3.0 | 2.0 | 112.0 | 10.0 | 10.5 | 39.0 | 40.2 |
| 10 | 33.0 | 28.3 | 6.7 | 40.0 | 3.0 | 5.0 | 208.0 | 20.0 | 19.0 | 30.0 | 19.3 |
| 11 | 44.0 | 25.8 | 6.7 | 40.0 | 4.1 | 5.0 | 132.0 | 22.0 | 8.2 | 46.0 | 13.4 |
| 12 | 16.7 | 50.0 | 8.3 | 40.0 | 5.0 | 5.1 | 112.0 | 10.0 | 10.5 | 39.0 | 24.0 |

### D.2 Analysis

In this section, we evaluate the total effect of $X_1$ on $Y$ because similar observations can be derived regarding other treatment variables. In this case study, we assume that $\{X_2, X_7\}$ is a subset of intermediate variables selected according to prior causal knowledge and that $\{X_3, X_8, X_{10}\}$ is a subset of covariates selected according to prior causal knowledge. In contrast, $\{X_4, X_5, X_6, X_9\}$ is a subset for which it is uncertain which element should be selected to evaluate the total effects.

For LASSO, adaptive LASSO, Elastic Net and PAL$_1$MA, $\{X_1, X_3, X_4, X_5, X_6, X_8, X_9, X_{10}\}$ were included as explanatory variables. In particular, the regression coefficients of $\{X_4, X_5, X_6, X_9\}$ are penalized in PAL$_1$MA. In contrast, regarding PCM Selector, $\{X_1, X_3, X_4, X_5, X_6, X_8, X_9, X_{10}\}$ were included as explanatory variables in the $L_p$-penalized loss function (8) with the response variables $\{X_2, X_7\}$, and the regression coefficients of $\{X_4, X_5, X_6, X_9\}$ were penalized. In addition, $\{X_1, X_2, X_3, X_4, X_5, X_6, X_7, X_8, X_9, X_{10}\}$ were included as explanatory variables in the $L_p$-penalized loss function (7) and the regression coefficients of $\{X_1, X_4, X_5, X_6, X_9\}$ were penalized. With respect to the TSLS method based on the front-door-like criterion, we used all intermediate variables and covariates to evaluate the total effect of $X_1$ on $Y$. For the OLS method based on the back-door criterion, we also used all covariates to evaluate the total effect of $X_1$ on $Y$. Furthermore, to characterize the estimation accuracy, the standard deviations were calculated based on the leave-one-out method.

Table G shows the results obtained by each regression analysis. Here, parameter tuning was conducted by the same procedure as in Section C.2. We also provide the violin plots of the estimated total effect by each regression analysis shown in Fig. E and the solution paths with the selected variables shown in Figs. F and G.

First, according to Okuno et al. (1986), the dilution ratio ($X_1$) is an important factor that increases both rust protection quality and visual appearance. However, from Table G and Fig. E, the total effect of $X_1$ on $Y$ is estimated as zero by LASSO, adaptive LASSO, and Elastic Net, which is problematic because it provides such a misleading interpretation that it is not useful to control $X_1$ to achieve the aim. In contrast, PAL$_1$MA, PCM Selector, the OLS method (with a back-door criterion based on all covariates), and the TSLS method (with the front-door-like criterion (not including $X_1$)) based on all covariates) estimate the total effect of $X_1$ on $Y$ as a negative value. Here, since $X_1$ is highly correlated with $X_2$ and $X_4$ $\left(\frac{\sigma_{x_1 x_2}}{\sqrt{\sigma_{x_1 x_1} \sigma_{x_2 x_2}}} = -0.593, \frac{\sigma_{x_1 x_4}}{\sqrt{\sigma_{x_1 x_1} \sigma_{x_4 x_4}}} = -0.686\right)$, compared to other correlation relationships between variables, the total effect of $X_1$ on $Y$ is estimated as a positive value and the estimation accuracy is worse when using the TSLS method with the front-door-like criterion including $X_1$.

Second, the OLS method provides an unbiased estimator of the total effect through a whole set of covariates that satisfy the back-door criterion, and the TSLS method also provides

an unbiased estimator of the total effect through a whole set of intermediate variables that satisfy the front-door-like criterion with a whole set of covariates that satisfy the back-door criterion. Given this finding, it is desirable for the estimators from the penalized regression analysis to be close to both the OLS estimate and the TSLS estimate. From this observation, from Table G and Fig. E, the estimates from PAL$_1$MA are close to the OLS estimates, but the estimates from PCM Selector are close to the TSLS estimates. The difference between the OLS (PAL$_1$MA) estimate and the TSLS (PCM Selector) estimate may be due to the small sample size problem or the model misspecification problem. In fact, Kuroki (2012) applied graphical modeling (Whittaker, 2009) based on some prior causal knowledge to the sample correlation matrix given by Okuno et al. (1986), and selected Fig. D by considering the simplicity ($dev = 34.28$, $df = 36$, $p$-value$= 0.55$). Here, note that both PCM Selector and PAL$_1$MA provide better estimation accuracy than the OLS and the TSLS methods. The standard deviation from PAL$_1$MA is lower than that from PCM Selector, but this difference seems not to be significant.

Third, from Figs. F and G, LASSO, adaptive LASSO, and Elastic Net select $X_8$, $\{X_5, X_6, X_8\}$, and $\{X_3, X_4, X_5, X_6, X_8\}$, respectively, which may be difficult to interpret the results from the viewpoint of causal inference because these sets of covariates do not satisfy the back-door criterion. In contrast, PAL$_1$MA selects $\{X_1, X_3, X_8, X_{10}\}$, which satisfies the back-door criterion. PCM Selector also selects $\{X_3, X_8, X_{10}\}$ regarding $\{X_2, X_7\}$; $\{X_2, X_7\}$ satisfies the front-door-like criterion relative to $(X_1, Y)$ with $\{X_3, X_8, X_{10}\}$. This implies that PAL$_1$MA and PCM Selector could help us to interpret the results from the viewpoint of causal inference.

Fourth, as shown in Figs. F and G, LASSO, adaptive LASSO, and Elastic Net estimate the total effect of $X_1$ on $Y$ as zero with zero standard deviation because $X_1$ is judged not to be active by these penalized regression analyses. In contrast, the estimated 95% confidence intervals from the OLS method, PAL$_1$MA, and PCM Selector do not include zero. From this observation, it is judged that $X_1$ has a negative effect on $Y$ by the OLS method, PAL$_1$MA, and PCM Selector, but the hypothesis that $X_1$ has no effect on $Y$ may not be rejected by LASSO, adaptive LASSO, Elastic Net, or the TSLS method. Here, it seems that PAL$_1$MA provides better estimation accuracy than PCM Selector. This seems to contradict Theorem 1, but it is not, because Theorem 1 is derived under the assumption that PAL$_1$MA and PCM Selector utilize the same set of covariates and the same weight matrix. In the case study, PAL$_1$MA selects the different sets of covariates and different weight matrices from the PCM Selector.

Figs. F and G show the solution paths for the penalty parameter when the other parameters are fixed at the values in Table G. PCM Selector and PAL$_1$MA automatically excluded $X_4, X_5, X_6$, and $X_9$, and it is uncertain whether they should be included given the value of the penalty parameter based on cross-validation. However, since cross-validation with datasets split into training and test datasets aims to achieve better prediction accuracy for the response variable and not proper qualitative variable selection, if we prefer to achieve proper qualitative variable selection from a causal inference perspective, the value of the penalty parameter can be selected according to the importance levels of the variables presented by the solution paths. Therefore, the development of optimal parameter tuning for PCM Selector is left for future work.

Table G. Results.

| Method | Estimate | SD | Parameters | | | | | |
|---|---|---|---|---|---|---|---|---|
| | | | $\lambda_1$ | $\rho_1$ | $\eta_1$ | $\zeta_1$ | $\xi_1$ | $\phi$ |
| LASSO | 0.000 | 0.000 | 0.416 | - | - | - | - | - |
| adaptive LASSO | 0.000 | 0.000 | 0.308 | - | 0.500 | - | - | - |
| Elastic Net | 0.000 | 0.030 | 0.416 | - | - | - | - | 0.340 |
| PAL$_1$MA | -0.250 | 0.089 | 0.252 | - | 0.900 | - | - | - |
| PCM Selector | -0.160 | 0.117 | 0.366 | 0.062 | - | 0.430 | 0.000 | |
| Front-door-like (including $x$) | 8.124 | 0.827 | - | - | - | - | - | - |
| Front-door-like (not including $x$) | -0.167 | 0.253 | - | - | - | - | - | - |
| Back-door | -0.268 | 0.249 | - | - | - | - | - | - |

Estimate: estimates of the total effect with $n = 12$; SD: standard deviation based on the leave-one-out method; $\lambda_1$: penalty parameter for $L_1$ penalization for the response model; $\rho_1$: penalty parameter for $L_1$ penalization for the mediator model; $\eta_1$, $\zeta_1$, $\xi_1$: tuning parameters; $\phi$: mixing parameter. The tuning parameters for weight vectors are adaptive LASSO: $\lambda' = 13.960$; PAL$_1$MA: $\lambda = 99.800$; PCM Selector: $\lambda = 120.105$, $\rho = 0.165$. All tuning parameters for bias correction were set to 0.

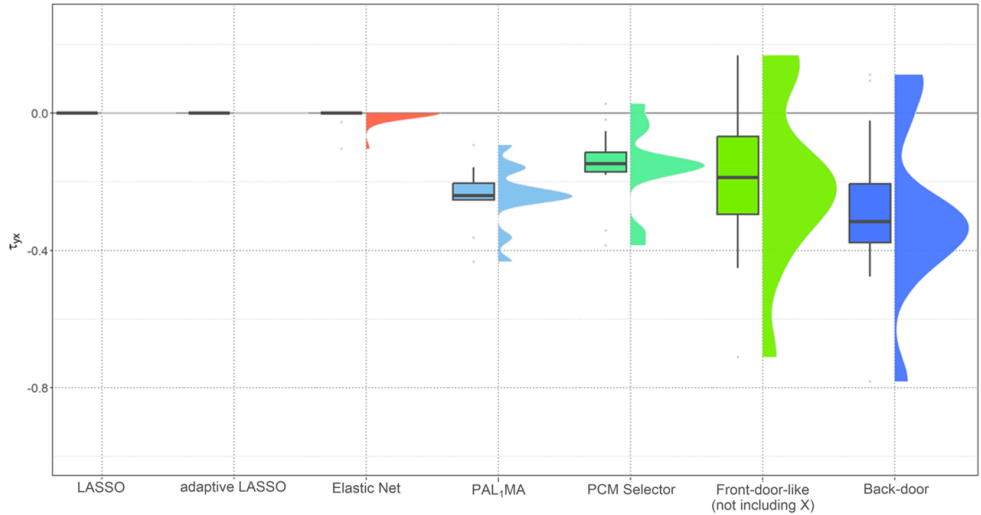

Fig. E. Violin plots of the case study for setting up the coating conditions for car bodies.

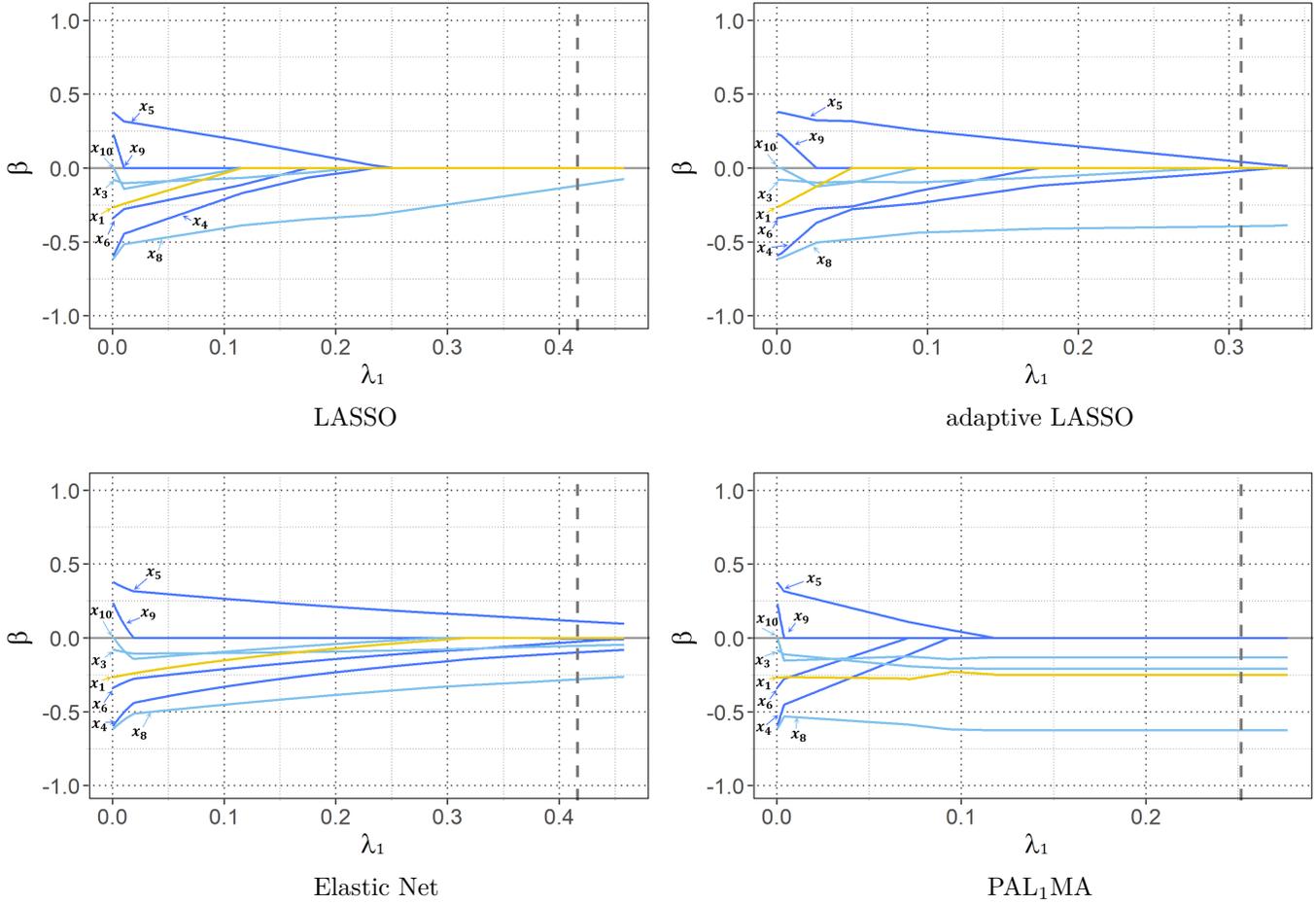

Fig. F. Solution paths for the penalty parameter $\lambda_1$ when the other parameters are fixed to the values in Table G. The dashed vertical line represents the values of $\lambda_1$ from Table G. The yellow line indicates the regression coefficient of $X_1$; the light blue line indicates the regression coefficients of covariates $\{X_3, X_8, X_{10}\}$; and the blue line indicates the regression coefficients of covariates $\{X_4, X_5, X_6, X_9\}$.

19

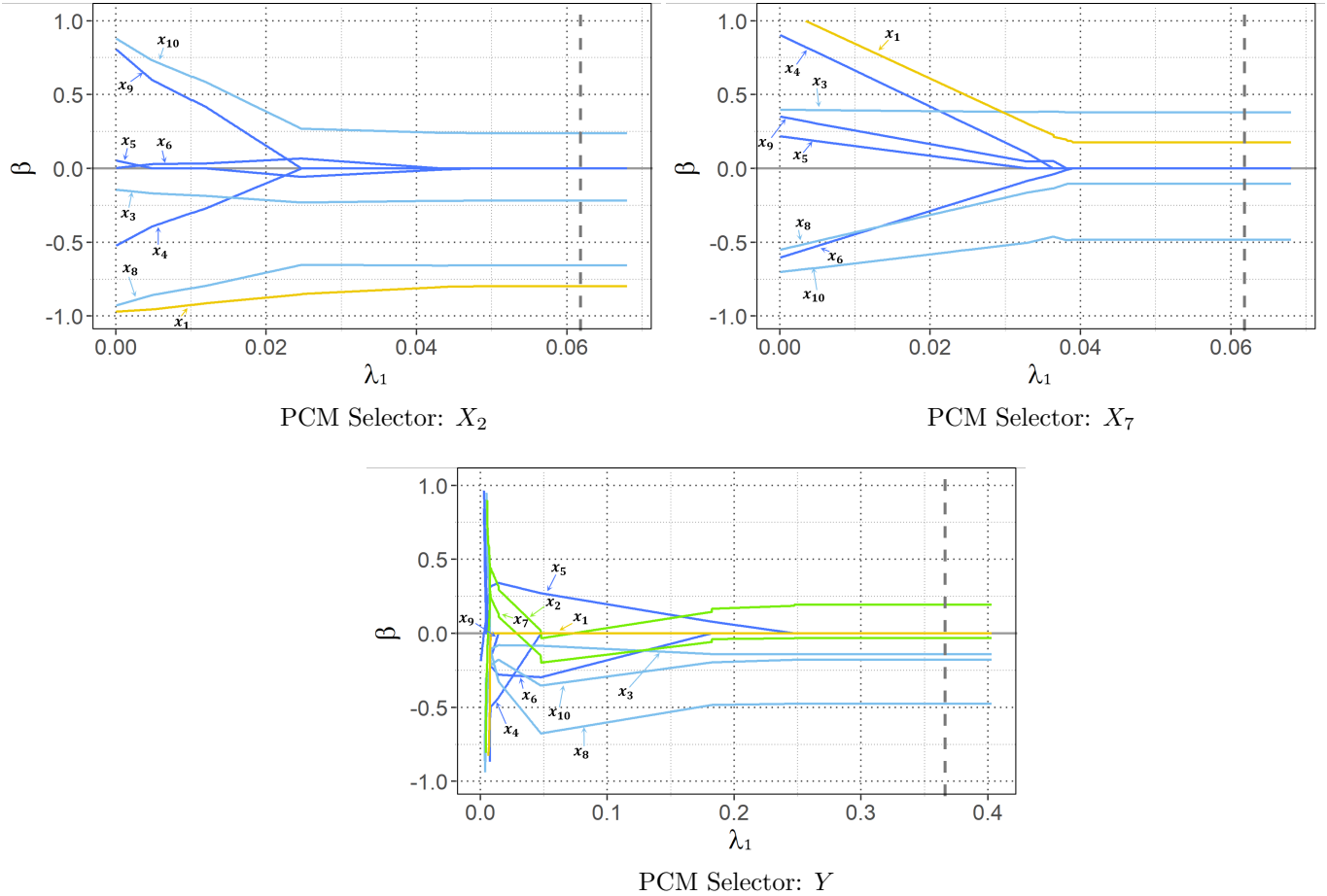

Fig. G. Solution paths for the penalty parameter $\lambda_1$ when the other parameters are fixed to the values in Table G. "PCM Selector: $X_2$" shows the regression coefficients with the response variable $X_2$ at the first stage, "PCM Selector: $X_7$" shows the regression coefficients with the response variable $X_7$ at the first stage, and "PCM Selector: $Y$" shows the regression coefficients with the response variable $Y$ at the second stage. The dashed vertical line represents the value of $\lambda_1$ from Table G. The yellow line indicates the regression coefficient of $X_1$; the light blue line indicates the regression coefficients of covariates $\{X_3, X_8, X_{10}\}$; the blue line indicates the regression coefficients of covariates $\{X_4, X_5, X_6, X_9\}$; and the green line indicates the regression coefficients of intermediate variables $\{X_2, X_7\}$.


\end{document}